\documentclass[prd,superscriptaddress,nofootinbib,twocolumn]{revtex4}
\usepackage{graphicx,natbib}
\usepackage{url}
\usepackage{color}
\usepackage{rotating}
\usepackage{amssymb,amsmath}

\newcommand{\Mobs}{M_{\rm obs}}
\newcommand{\Mbias}{M_{\rm bias}}
\newcommand{\Mth}{M_{\rm th}}
\newcommand{\siglnM}{\sigma_{\ln M}}
\newcommand{\zphot}{z^{\rm p}}
\newcommand{\zbias}{z^{\rm bias}}
\newcommand{\sigz}{\sigma_{z}}
\newcommand{\Msun}{M_{\odot}}
\newcommand{\DE}{\Omega_{\rm DE}}

\newcommand{\sigp}{\sigma_{\rm prior}}
\newcommand{\sigf}{\sigp^{\rm MF/B}}
\newcommand{\sigo}{\sigp^{\Mobs}}

\begin{document}

\title{Sensitivity of galaxy cluster dark energy constraints to halo modeling uncertainties}
\author{Carlos E. Cunha, August E. Evrard}
\email{ccunha@umich.edu}
\affiliation{%
Department of Physics, University of Michigan, Ann Arbor, Michigan 48109
}

\date{\today}

\begin{abstract}

We perform a sensitivity study of dark energy constraints from galaxy cluster surveys to 
uncertainties in the halo mass function, bias and the mass--observable relation.  
For a set of idealized surveys, we evaluate cosmological constraints as 
priors on sixteen nuisance parameters in the halo modeling are varied.   
We find that surveys with a higher mass limit are more sensitive to mass--observable uncertainties 
while surveys with low mass limits that probe more of the mass function shape and evolution are more 
sensitive to mass function errors.   
We examine the correlations among nuisance and cosmological parameters.  
Mass function parameters are strongly positively (negatively) correlated with $\DE$ ($w$).  
For the mass--observable parameters, $\DE$ is most sensitive
to the normalization and its redshift evolution while $w$ is more sensitive to 
redshift evolution in the variance.  
While survey performance is limited mainly by mass--observable uncertainties, 
the current level of mass function error is responsible for up to a factor of two degradation
in ideal cosmological constraints.   
For surveys that probe to low masses ($10^{13.5} h^{-1} \Msun$), even percent-level 
constraints on model nuisance parameters result in a degradation of 
$\sim \sqrt{2}$ (2) on $\DE$ ($w$) relative to perfect knowledge.
\end{abstract}

\maketitle
\section{Introduction}\label{sec:intro}

The spatial abundance of galaxy clusters is a potentially powerful 
approach to test the nature of dark energy \citep{cun09,sah09,voi05,bat03,ros02,hai01,mar06}.   
As is the case for all dark energy probes, modeling and other systematic 
uncertainties in the data analysis must be identified and controlled.  
The primary uncertainties for cluster counts involve theoretical 
modeling of the halo space density as a function of mass --- the {\sl mass function} --- 
along with uncertainties associated 
with modeling the halo mass selection function of a specific survey.  

While recent studies have derived consistent and competitive 
cosmological constraints using X-ray \cite{man08,vik09,hen09} and 
optical \cite{gla07,roz09} cluster surveys, the larger volume and 
improved sensitivity of upcoming surveys motivate a stronger 
effort to study sources of systematic error.   
Early work on forecasting cluster constraints focused attention 
on systematics arising from the mass--observable relation 
(\cite{lev02,maj03,maj04}, etc.).  
These errors fall into two categories:  i) bias arising from uncertain 
knowledge of the normalization and ii) mass variance at fixed observable 
signal or, equivalently,  scatter in the mass-signal relation.  
The redshift evolution of these effects are particularly 
important \cite{lim04,lim05,lim07,cun08}.  
If the scatter is large, the functional form of the 
mass--observable likelihood becomes important \citep{shaw09}.

Uncertainties in the mass--observable relation can be characterized 
with a combination of empirical and theoretical approaches.  
For example, the slope and normalization of the intracluster 
gas thermal energy can be probed with X-ray and 
Sunyaev-Zel'dovich (SZ) observations \citep{arnaud05, maughan06, 
vikhlinin06, morandi07, bonamente08, zhang08, pratt09} 
while the variance and evolution with redshift can be predicted 
in a model-dependent manner using  multi-fluid simulations 
\citep{bialek01, borgani04, dasilva04, kravtsov06, ascasibar06, muanwong06, puchwein08, aghanim09, stanek09b}.  
For example, a `preheated' gas treatment that matches the low-redshift 
X-ray luminosity--temperature relation predicts weakly increasing 
scatter in the integrated thermal SZ signal at fixed mass, 
from $13\%$ at  $z=0$ to $18\%$ at $z=1$ \citep{stanek09b}.  

Theoretical uncertainty in the mass function has not been extensively 
studied in previous Fisher forecasts.   
This is partly because mass--observable uncertainties are typically 
dominant in comparison, and partly because mass function 
calibrations by the simulation community have been evolving.   

The original mass function of Press \& Schechter \cite{pre74}, 
derived with support of 1000-particle N-body simulations,  has evolved into a 
number of forms with billion-particle simulation support, among them the Sheth-Tormen \cite{she99}, 
Jenkins \cite{jen01}, Evrard \cite{evr02}, Warren \cite{war06} 
and Tinker \cite{tin08} mass functions.   The Warren parameterization has received a recent update 
based on an ensemble of large-volume 
simulations\footnote{The MICE simulations, see http://www.ice.cat/mice.}.  
Correction for systematics associated with initial condition generation and other effects 
lead to an upward revision in the mass function of $\gtrsim10$ percent at mass scales probed 
by SZ surveys \citep{cro09}.  

Different measures of mass, based on either particle percolation or 
spherical filtering, are employed by the aforementioned studies 
(see \cite{laceyCole96,white02,lukic09} for discussion of these mass measures), but  all 
use the filtered linear power spectrum, $\sigma(M)$, as a similarity 
variable for expressing halo counts and clustering.  
While the Tinker calibration \cite{tin08} improved the statistical accuracy 
of the mass function to the 5\% level in number density, the effects of gas dynamics, absent in that work, 
may produce deviations that are larger than this \cite{stanek09a}.   We consider $\sim 10\%$ in space 
density as a reasonable estimate of the current level of uncertainty in the mass function at cluster scales.  

The cited mass function calibrations assume a standard 
$\Lambda$CDM cosmology, with a Gaussian initial density field evolved 
under general relativity with a cosmological constant.   
Extensions to non-standard assumptions are emerging (see e.g. \cite{lov08,dal08,gro09}), 
but we do not consider them here.

In this paper, we extend previous Fisher studies of dark energy 
constraints by including theoretical uncertainties in the halo 
mass function and clustering bias and by employing a more generous 
treatment of bias and scatter evolution in the mass--observable relation.  
We frame the analysis in terms of two idealized surveys roughly patterned after 
the South Pole Telescope\footnote{{\tt http://pole.uchicago.edu/}} (SPT) SZ survey 
and the Dark Energy Survey\footnote{{\tt https://www.darkenergysurvey.org/}} (DES) 
optical survey.  Our baseline idealizations, which  assume perfect halo selection 
(100\% completeness) above a redshift-independent mass threshold and  constant 
mass--observable scatter, provide a simple, rational basis upon which complexities, 
in the form of nuisance parameters discussed below, can be added.  
The ``SZ-motivated'' survey has a high mass threshold and small mass--observable scatter while 
the ``optical-motivated'' case has a lower mass threshold and larger mass scatter.   
We refer to these simply as `SZ' and `optical' for the remainder of the paper.

Our main aim is to study performance  degradation in $\DE$ and $w$ constraints 
as a function of prior parameter knowledge.  We also provide a first look at 
parameter correlations for the full model, finding them (unsurprisingly) complex.
While writing up this work, we learned of a similar study by Wu, Zentner \& Wechsler \cite{wu09b}, who address the effect of mass function systematics on the dark energy figure of merit from cluster surveys.   Their approach is complementary to ours, in that they adopt a set of independent nuisance parameters to model systematic error in  binned representations of the mass function and halo bias.  While this non-parametric approach allows them to more easily identify regions of mass and redshift space that most sensitively affect dark energy constraints, a disadvantage is that it disregards correlations between different mass/redshift bins.  Wu et al.~also explore a time-dependent equation of state, $w(a)$, and express their results in terms of a figure of merit degradation.   They do not emphasize the interplay between mass--observable and mass function systematic errors, as we do here.

The paper is organized as follows.  In Sec. \ref{sec:counts} we 
briefly review the formalism for extracting dark energy constraints
from cluster counts and variance in counts and present our 
parameterization of the mass--observable relations, the mass function 
and galaxy bias.
We present results in Sec. \ref{sec:res}, offer a critique in Sec. \ref{sec:disc} and conclude in Sec. \ref{sec:conc}.

\section{DE from cluster counts and clustering}\label{sec:counts}

The subject of deriving cosmological constraints from cluster number 
counts and clustering of clusters has been treated extensively in 
the literature (see e.g. \cite{cun08,lim04,lim05,lim07}).
In this section we follow closely the approach described in \cite{cun08}.

The comoving number density of clusters at a given redshift $z$ with observable in
the range $\Mobs^{\alpha} \leq \Mobs \leq \Mobs^{\alpha+1}$ is given by

\begin{eqnarray}
\bar n_{\alpha}(z) &\equiv& \int_{\Mobs^{\alpha}}^{\Mobs^{\alpha+1}} \frac{d \Mobs}{\Mobs}
\int {\frac{dM}{M}} { \frac{d \bar n}{d\ln M}}
p(\Mobs | M)
\label{eqn:nofz}
\end{eqnarray}

Uncertainties in the redshifts distort the volume element.
Assuming photometric techniques are used to determine the redshifts of 
the clusters (hereafter photo-z's) and a perfect angular selection the mean number of clusters 
in a photo-z bin $\zphot_i \le \zphot \le \zphot_{i+1}$ is

\begin{eqnarray}
\bar m_{\alpha,i} &=& \int_{\zphot_i}^{\zphot_{i+1}} d\zphot 
\int dz \frac{dV}{dz}  \bar n_{\alpha} W_{i}^{\rm th}(\Omega) p(\zphot| z)
\label{eqn:numwin}
\end{eqnarray}

\noindent where $W^{\rm th}_{i}(\Omega)$ is an angular top hat window function.
We parameterize the probability of measuring a photometric
redshift, $\zphot$, given the true cluster redshift $z$ as \cite{lim07}  
\begin{eqnarray}
p(\zphot| z) &=& \frac{1}{\sqrt{2\pi} \sigz} \exp\left[ -y^2(\zphot) \right]
\label{eqn:pzpzs}
\end{eqnarray}

\noindent where

\begin{eqnarray}
y(\zphot)&\equiv& \frac{\zphot -z -\zbias}{ \sqrt{2}\sigz}, 
\end{eqnarray}

\noindent $\zbias$ is the photometric redshift bias and $\sigz$ is the scatter in the 
photo-z's. 
We fix the photo-z bias and scatter at $0$ and $0.02$ throughout this paper.


The sample covariance of counts $m_{\alpha,i}$ is, given by \citep{hu03}

\begin{eqnarray}
S_{ij} &=&\langle (m_{\alpha,i} -\bar m_{\alpha,i})(m_{\alpha,j} - \bar m_{\alpha,j})\rangle \label{eqn:sija} \\
&=& b_{\alpha,i} \bar m_{\alpha,i} b_{\alpha,j} \bar m_{\alpha,j} \nonumber \\
&&\times \int{\frac{d^3 k}{(2\pi)^3}} W_i^*({\bf k})W_j({\bf k})\sqrt{P_i(k)P_j(k)}, \label{eqn:sijb}
\end{eqnarray}

\noindent where $b_{\alpha,i}(z)$ is the 
average cluster linear bias defined as
\begin{eqnarray}
b_{\alpha,i}(z) &=& \frac{1}{\bar n_{\alpha,i}(z)}  \int_{\Mobs^{\alpha}}^{\Mobs^{\alpha+1}} \frac{d{\Mobs}}{\Mobs}\int \frac{d M}{M} \nonumber \\
&&\times \frac{d \bar n_{\alpha,i}(z)}{d\ln M} b(M;z)p(\Mobs|M).
\label{eqn:bias}
\end{eqnarray}

\noindent $W_i^*({\bf k})$ is the Fourier transform of the top-hat 
window function and $P_i(k)$ is the linear power spectrum at the centroid 
of redshift bin $i$.
We present our choice for $b(M;z)$ in \ref{sec:mfbpar}, when we discuss the
parameterization of the errors in the mass function and galaxy bias.
We only calculate covariance terms for which $i=j$ since off-diagonal 
terms are negligible.

Following \cite{lim07}, we find that the window function  $W_i^*({\bf k})$ 
in the presence of photo-z errors is given by

\begin{eqnarray}
W_i({\bf k})=&&2\exp{\left[ i k_{\parallel} \left( r_i+\frac{ {\zbias_i} }{ {H_i} } \right) \right] }
              \exp{ \left[- \frac{{\sigma_{z,i}^2 k_{\parallel}^2}}{{2H_i^2}}  \right] } \nonumber \\
           && \times  \frac{\sin( k_{\parallel} \delta r_i/2) }{k_{\parallel} \delta r_i/2} 
              \frac{J_1(k_{\perp} r_i \theta_s)}{k_{\perp} r_i \theta_s}. 
\label{eqn:window}
\end{eqnarray}
\noindent Here $r_i=r(\zphot_i)$ is the angular diameter distance to the 
$i^{\rm th}$ photo-z bin, and $\delta r_i=r(\zphot_{i+1})-r(\zphot_{i})$.
Similarly, $H_i=H(\zphot_i)=H(z)$, $\zbias_i=\zbias(\zphot_i)=\zbias(z)$, and 
$\sigma_{z,i}=\sigma_z(\zphot_i)=\sigma_z(z)$.
We assumed that $H(z)$, $\zbias(z)$, and $\sigma_z(z)$ are constant inside each bin.
The variables $k_{\parallel}$ and $k_{\perp}$ represent parallel and perpendicular 
components of the wavenumber ${\bf k}$ relative to the line of sight.

Define the covariance matrix of halo counts
\begin{equation}
C_{ij} = S_{ij} + {\bar m_i} \delta_{ij}
\label{eqn:covdat}
\end{equation}
\noindent where ${\bar m_i}$ is the vector of mean counts defined in Eq. (\ref{eqn:numwin}) and 
$S_{ij}$ is the sample covariance defined in Eq. (\ref{eqn:sijb}).
The indices $i$ and $j$ here run over all mass and redshift bins.
Assuming Poisson noise and sample variance are the only sources of 
noise, the Fisher matrix is, \citep{hu06,lim04,hol01}

\begin{equation}
F_{\alpha\beta}=  \bar{\bf m}^t_{,\alpha} {\bf C}^{-1}
 \bar{{\bf m}}_{,\beta} 
+ \frac{1}{2} {\rm Tr} [{\bf C}^{-1} {\bf S}_{,\alpha}
 {{\bf C}}^{-1} {\bf S}_{,\beta} ],
 \label{eqn:fishmat}
\end{equation}
where the ``,'' denote derivatives with respect to the model parameters.
The first term on the right-hand side contains the ``information'' 
from the mean counts, $\bar m$.
The $S_{ij}$ matrix only contributes noise to this term, and 
hence only reduces its information content.
The second term contains the information from the sample covariance.

\subsection{Systematics in the mass--observable relation}\label{sec:nupar}

We introduce six degrees of freedom in the sector of the model that links 
halo mass to an observable signal.
We assume a log-normal form for the probability of measuring an observable signal, denoted $\Mobs$, given 
true mass $M$, 

\begin{equation}
p(\Mobs | M) = \frac{1}{ \sqrt{2\pi} \siglnM}  \exp\left[ -x^2(\Mobs) \right] ,
\label{eqn:mobsDefn}
\end{equation}

\noindent where

\begin{equation}
x(\Mobs) \equiv \frac{ \ln \Mobs - \ln M - \ln  \Mbias(\Mobs,z)}{ \sqrt{2} \siglnM(\Mobs,z)}.
\label{eqn:x1mobs}
\end{equation}

We model systematic error in the mass proxy by introducing a redshift-dependent bias and variance
\begin{eqnarray}
{\rm ln}\Mbias(z)&=&B_0 + B_1(1+z), \label{eqn:mbiasdefsz}\\
\siglnM^2(z)&=&\sigma_{0}^2 + \sum_{i=1}^{3}S_iz^i \label{eqn:msigdefsz}, 
\end{eqnarray}
where $B_0$, $B_1$, $\sigma_0$ and the three variance coefficients are assumed to be independent of mass.    
Our default assumption is that $\sigma_0$ is non-zero, with values discussed  in 
Sec. \ref{sec:clustermass}, while the remaining parameters have a fiducial value of zero.   
All six parameters are taken as degrees of freedom and varied in the Fisher 
analysis presented in Sec. \ref{sec:res}.

\subsection{Systematics in the halo mass function and bias}\label{sec:mfbpar}

In the sector of the model describing the halo mass function, we add ten more degrees of freedom.  
We write the space density of halos as 
\begin{eqnarray}
\frac{dn}{dM}=f(\sigma)\frac{\bar \rho_m}{M}\frac{d\ln \sigma^{-1}}{dM} \label{eqn:mfunc}
\end{eqnarray}

\noindent and adopt the Tinker parameterization of $f(\sigma)$ \citep{tin08}
\begin{eqnarray}
f(\sigma)=A\left[\left(\frac{\sigma}{b}\right)^{-a} +1 \right]e^{-c/{\sigma^2}}. \label{eqn:tinker}
\end{eqnarray}
\noindent Following \cite{tin08}, we allow the first three parameters of $f(\sigma)$ 
to vary with redshift, so that
\begin{eqnarray}
A(z)=A_0(1+z)^{A_x} \\
a(z)=a_0(1+z)^{a_x} \\
b(z)=b_0(1+z)^{-\alpha}
\end{eqnarray}

For fiducial parameters, we adopt the values of \cite{tin08} at $\Delta=200$:
$A_0 = 0.186$, $A_x =-0.14$, $a_0 = 1.47$, $a_x = -0.06$, $b_0 = 2.57$, $\log_{10}(\alpha) = {(\frac{0.75}{log(\Delta/75)})^{1.2}}$, and $c = 1.19$.
As Tinker et al. \cite{tin08} explain, A controls the overall amplitude of $f(\sigma)$, 
$a$ controls the tilt, and $b$ sets the mass scale where the power law in $f(\sigma)$
becomes significant.

We adopt the $b(M,z)$ fit of \cite{she99} for the galaxy bias
\begin{equation}
b(M,z) = 1 + \frac{a_c \delta_c^2/\sigma^2 -1}{\delta_c} 
         + \frac{ 2 p_c}{\delta_c [ 1 + (a \delta_c^2/\sigma^2)^{p_c}]} , 
\label{eqn:biasofmz}
\end{equation}
\noindent and choose the fiducial values for the parameters to be
$a_c=0.75$, $\delta_c=1.69$, and $p_c= 0.3$.  

In total, the mass function and bias introduce ten additional parameters, 
and we consider all of these as  degrees of freedom in the Fisher analysis 
presented in \S~\ref{sec:res}.  The assumption that the bias is independent of the mass function
is very conservative.  Manera et al. \cite{man09} show that, in the range of scales we are interested
in, the bias can be predicted to roughly $\sim 10\%$ accuracy given the mass function. 

\subsection{Fiducial Parameter Values}\label{sec:fiducial}

The fiducial values of the sixteen nuisance parameters are summarized in Table \ref{tbl:params}.   These parameters control the underlying counts and clustering of equations~(\ref{eqn:nofz}) and (\ref{eqn:bias}) via the explicit forms of equations (\ref{eqn:mobsDefn}), (\ref{eqn:mfunc}) and (\ref{eqn:biasofmz}), and thereby produce the sample covariance, equation~(\ref{eqn:covdat}), and Fisher matrix, equation~(\ref{eqn:fishmat}).

\begin{table}
\caption{Halo modeling nuisance parameters }
\begin{center}
\leavevmode
\begin{tabular}{ l c  r  r  }\hline \hline
\multicolumn{1}{c}{Class} & \multicolumn{1}{c}{Name} & \multicolumn{1}{c}{Fid. Value} & \multicolumn{1}{c}{Notes} \\
\hline
&$B_0$&0.0&constant bias\\
&$B_1$&0.0&z-dependent bias\\
$\Mobs$&$\sigma_0^2$&0.04,0.625&constant variance\\
&$S_1$&0.0&$z^1$ variance \\
&$S_2$&0.0&$z^2$ variance\\
&$S_3$&0.0&$z^3$ variance\\
\hline
&$A_0$&0.186&z-independent\\
&$a_0$&0.147&  '' \\
Mass&$b_0$&2.57& ''\\
Function&$c$&1.19& ''\\
&$A_x$&-0.14&$A(z) = A_0 (1+z)^{A_x}$\\
&$a_x$&-0.06& $a(z)=a_0(1+z)^{a_x}$\\
&$\alpha$&0.0107&$b(z) = b_0 (1+z)^\alpha$\\
\hline 
&$a_c $&0.75&z-independent\\\
Bias&$\delta_c $&1.69&''\\
&$p_c $&0.30&''\\
\hline
\label{tbl:params}
\end{tabular}
\end{center} 
\end{table}

\subsection{Reference Model Surveys}\label{sec:clustermass}

We apply our tests to four distinct surveys consisting of a fiducial and three options. 
All are assumed to cover a sky area of 4000 square degrees and extend to a 
limiting redshift $z_{\rm max} =2.0$.  
The test surveys differ only in two parameters: the mass threshold, $M_{\rm th}$,  and the zero--redshift  
variance in the mass--observable relation, $\sigma_0^2$.  

Our chosen survey parameters, given in Table \ref{tbl:constraints}, 
represent capabilities likely to be realized in the near future using 
sub-mm and optical/NIR observations.  
For example, the South Pole Telescope  (SPT) is expected to detect 
clusters above $\Mth = 10^{14.2} h^{-1} \Msun$  up to a redshift of 2 
(see e.g. \cite{car02}), with photometric redshifts available from 
DES+VISTA\footnote{\tt http://www.vista.ac.uk/} imaging, 
the Blanco Cosmology Survey, and the Magellan Telescope.  
DES+VISTA will have internal capability to detect clusters optically 
with techniques known to work above $\Mth = 10^{13.5} h^{-1} \Msun$ 
(see e.g. \cite{koe07} and \cite{joh07}).

We subdivide the sky into 400 bins  of 10 sq. degrees each, and calculate 
the counts and sample variance using mass bins of width 
$\log(\Delta \Mobs)=0.2$ with the exception of the highest mass bin, 
which we extend to infinity.
We set the width of our redshift bins to $\Delta \zphot =0.1$.
These bin sizes imply 20 redshift bins and 5 bins of mass for the 
surveys 1 and 2.
For Surveys 3 and 4, we divide the mass range 
$10^{13.5}\leq M_{obs}^{\rm opt} \leq 10^{14.2} h^{-1} \Msun$ 
into 5 bins and use the same mass binning as the Surveys 1 and 2 for 
$M_{obs}^{\rm opt} > 10^{14.2} h^{-1} \Msun$, 
with a total of 10 mass bins and 20 redshift bins.

We assume fiducial cosmological parameters based on the fifth year data release
of the Wilkinson Microwave Anisotropy Probe (WMAP5, \cite{kom08}).
Thus, we set the baryon density, $\Omega_b h^2=0.0227$,
the dark matter density, $\Omega_m h^2 =0.1326$, the normalization of the power 
spectrum at $k=0.05 {\rm Mpc}^{-1}$, $A_{s}=4.625 \times 10^{-5}$, the
tilt, $n=0.963$, the optical depth to reionization, $\tau=0.087$, the 
dark energy density, 
$\Omega_{\rm DE}=0.742$, and the dark energy equation of state, $w=-1$.
In this cosmology, $\sigma_8=0.796$.

With the exception of $w$, the cosmological parameters we used have 
been determined to 
an accuracy of a few percent.
Extrapolating into the future, we assume $1\%$ priors on all 
cosmological parameters 
except $\Omega_{\rm DE}$ and $w$.
We used CMBfast \citep{sel96}, version 4.5.1, to calculate the 
transfer functions.
We do not explore time evolution of $w$ in this work.  In cosmologies with a time varying equation of state, $w(a)$, the relationship between halo model parameters and the error in principal components of $w(a)$ would differ from that reported here, with redshift terms having a larger impact.   An important aim of next generation surveys is to test $w \ne -1$, the degree to which dark energy differs from vacuum energy.

\section{Results}\label{sec:res}

\begin{table*}
\caption{Surveys parameters and constraints on cosmological parameters }
\begin{center}
\leavevmode
\begin{tabular}{ l c r r r r r r }\hline \hline
\multicolumn{4}{c}{} & \multicolumn{2}{c}{Sharp priors} & \multicolumn{2}{c}{No priors} \\
Survey & $\Mth [h^{-1} \Msun]$& $\sigma_0$ & $N_{\rm tot}$ & $\sigma(\DE)$ & $\sigma(w)$ & $\sigma(\DE)$ & $\sigma(w)$ \\
\hline
Fid. & $10^{14.2}$ & 0.2 & 8,400  & 0.010  & 0.050 & 0.91 & 2.19 \\
1 & $10^{14.2}$& 0.5 & 16,400  & 0.0083  & 0.039 & 0.82  & 1.81 \\
2 & $10^{13.5}$& 0.2 & 359,600  & 0.0025  & 0.011 & 0.098  & 0.23 \\
3 & $10^{13.5}$ & 0.5 & 482,400  & 0.0023  & 0.0097 & 0.22  & 0.35 \\
\hline \hline
\label{tbl:constraints}
\end{tabular}
\end{center} 
\end{table*}

The baseline, absolute accuracy in $\Omega_{\rm DE}$ and $w$ measurements 
from the four test surveys are given in Table \ref{tbl:constraints}.   
The sharp priors columns represent the ideal condition of perfect knowledge of the 
mass function and mass--observable relation, {\sl i.e.}, delta function priors
on all the mass function and mass--observable relation parameters.   
The no priors columns give results assuming a high degree of ignorance in 
parameter values.  

From Table \ref{tbl:constraints}, we see that a factor $10^{0.7}$
decrease in the mass threshold, $\Mth$, improves the constraints on both $\DE$ 
and $w$ by a factor of $\sim 4$.
The improvement results from the increase in cluster counts as well
as an increase in the exposed range of the mass function.
Surveys 2 and 3 find a factor of 27 and 18 more clusters than the fiducial survey
and survey 1, respectively. 
Increased scatter in $\Mobs$ results in an increase in counts because of 
the steepness of the mass function near the $\Mth$.
The negative mass function slope implies that more objects will be up-scattered 
from below $\Mth$ than down-scattered. 
If the scatter is perfectly known, the increase in counts yields better
cosmological constraints \citep{cun08,lim05}. 
This optimistic result must be interpreted with caution. 
A large mass--observable scatter may reflect poor selection, or 
contamination by projection or intrinsic sources.   
Projection is known to produce non-Gaussian features in the 
mass--observable relation \citep{coh06} that can bias constraints if not 
correctly accounted for (Erickson et al., in prep. \cite{smi09}).  

The last columns of Table \ref{tbl:constraints} demonstrate that large 
degradations in cosmological accuracy result from essentially complete 
ignorance of the halo modeling parameters.   We turn now to study the 
transition between the regimes of complete knowledge and complete 
ignorance by varying the prior uncertainties on the sixteen nuisance parameters.

\subsection{Degradation from Model Systematic Error}\label{sec:degrad}

For the parameters controlling the 
mass--observable relation ($\Mobs$) and the mass-function/bias (MF/B), 
we introduce prior uncertainties, $\sigma_{\rm prior}$, that represent 
errors from previous observation or simulation.  
Because the $\Mobs$ and MF/B parameters have different dimensions, we define 
the priors on them differently, so as to make the prior uncertainties more
directly comparable.

We define the prior uncertainty on $\Mobs$ nuisance parameters, 
$\sigo$, such that the prior $F^{ii}_{\rm prior}$ added to the $i^{\rm th}$ diagonal 
element of the Fisher matrix is
\begin{equation}
F^{ii}_{\rm prior}=\left (\frac{1}{\sigo}\right)^2. \label{eqn:primobs}
\end{equation}
\noindent The prior uncertainty on an MF/B nuisance parameter, $x_i$, 
is defined as
\begin{equation}
F^{ii}_{\rm prior}=\left (\frac{1}{x_i^2\sigf}\right)^2. \label{eqn:primf}
\end{equation}
\noindent With this definition, $\sigf$ corresponds to the prior fractional uncertainty
on each mass-function/bias nuisance parameter. 
The uncertainties in the $\Mobs$ nuisance  parameters are already fractional since the 
mass--observable relation is defined in terms of the logarithm of the mass.  

\begin{figure*}[t]
\vspace{-2.2truecm}
  \begin{minipage}[t]{85mm}
    \begin{center}
      \resizebox{100mm}{!}{\includegraphics[angle=0]{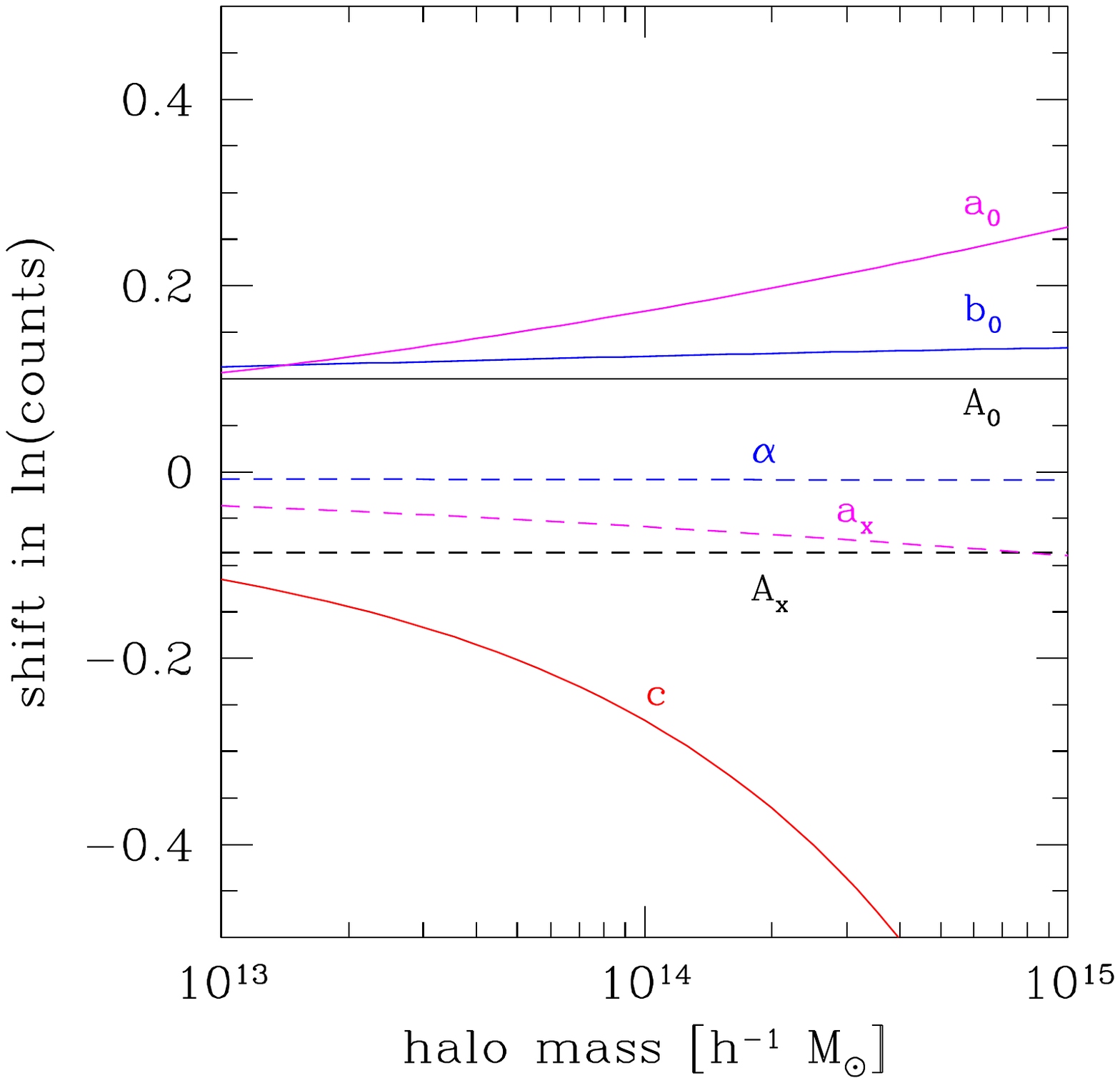}}
    \end{center}
  \end{minipage}
  \begin{minipage}[t]{85mm}
    \begin{center}
      \resizebox{100mm}{!}{\includegraphics[angle=0]{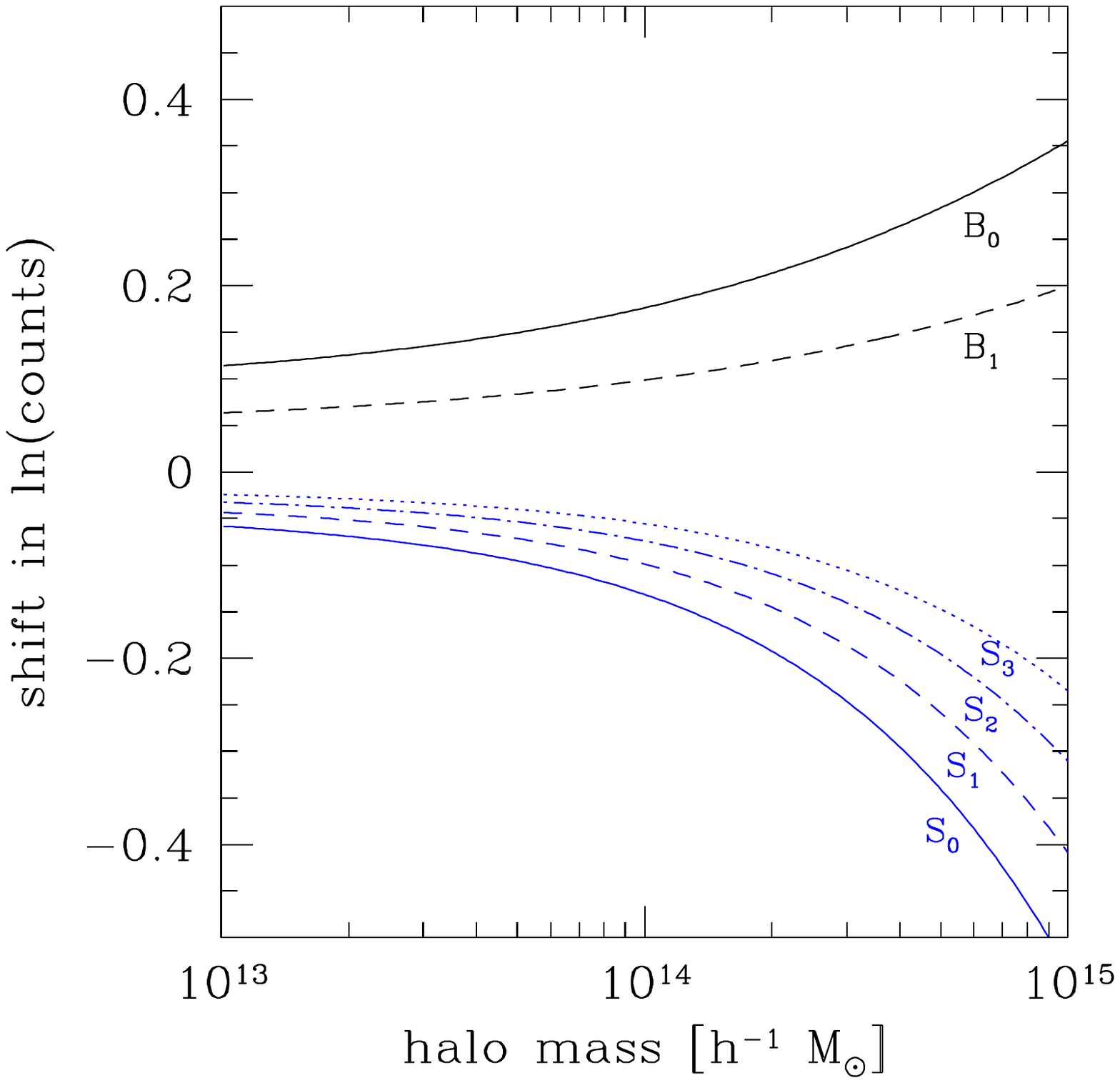}}
    \end{center}
  \end{minipage}
\vspace{-3.0truecm}
  \caption{The sensitivity of the mass function to variation in Tinker mass function (left panel) and mass--observable 
(right) parameters is illustrated for redshift $z=0.75$, roughly the median redshift of the surveys considered here.  
The change in the natural logarithm of number counts as a function of mass is shown while each of the labeled parameters 
is changed by a fractional amount, $\sigf=0.1$ (left panel) or a fixed amount, $\sigo = \pm 0.1$ (right panel).  
In the latter case, the bias terms are varied by $0.1$ and the variance terms by $-0.1$.  
 }\label{fig:tinkerParams}
  \end{figure*}

The {\it left plot} of Fig.~\ref{fig:tinkerParams} gives a sense of the magnitude of the shifts 
in number counts 
as each of the Tinker MF parameters are increased by a fractional amount, $\sigf=0.1$. 
We evaluate the mass function shifts for the fiducial cosmology at redshift $z = 0.75$, 
roughly the median redshift of our model surveys.   Solid lines show the effects of 
varying the constant terms while dashed lines vary the redshift-dependent factors.  
At $10^{14} h^{-1}\Msun$, the shift in number ranges from $+0.2$ (varying $a_0$) 
to $-0.3$ (varying $c$).  
The derivatives are positive for the constant terms, with the exception of 
$c$, while the derivatives with respect to the $z$-dependent terms are negative.   
From the figure, we see that a $10\%$ change in $\alpha$ causes very 
little change in the mass function. 
However, as we shall see in Sec. \ref{sec:correl}, $\alpha$ and the other redshift
evolution parameters ($a_x$ and $A_x$) are nearly perfectly 
correlated (anti-correlated) with $\DE$ (w), suggesting that the redshift evolution
of the mass function needs to be well understood to avoid degradations in cosmological
parameter constraints.
In the {\it right plot}, we show the dependence of the observed 
mass function on the $\Mobs$ nuisance parameters, at the same redshift as above. 
The impact of the $\Mobs$ nuisance parameters increase with mass because the slope of the 
mass function increases with mass.
The steeper the slope, the more significant is the imbalance between clusters up-scattered
or down-scattered into a given mass bin due to bias or scatter in $\Mobs$.

\begin{figure*}
  \begin{minipage}[t]{85mm}
    \begin{center}
      \resizebox{100mm}{!}{\includegraphics[angle=0]{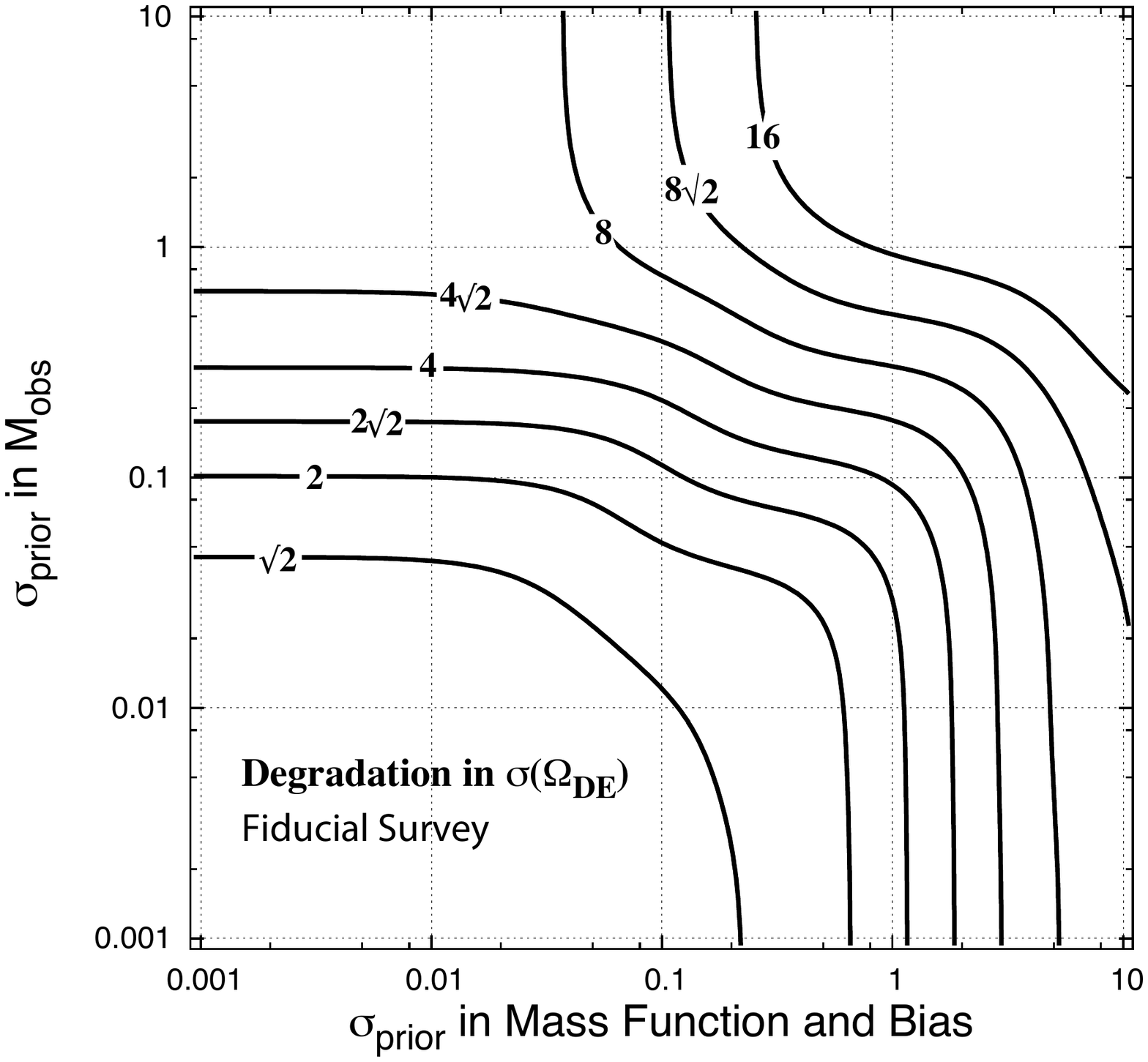}}
    \end{center}
  \end{minipage}
  \begin{minipage}[t]{85mm}
    \begin{center}
      \resizebox{100mm}{!}{\includegraphics[angle=0]{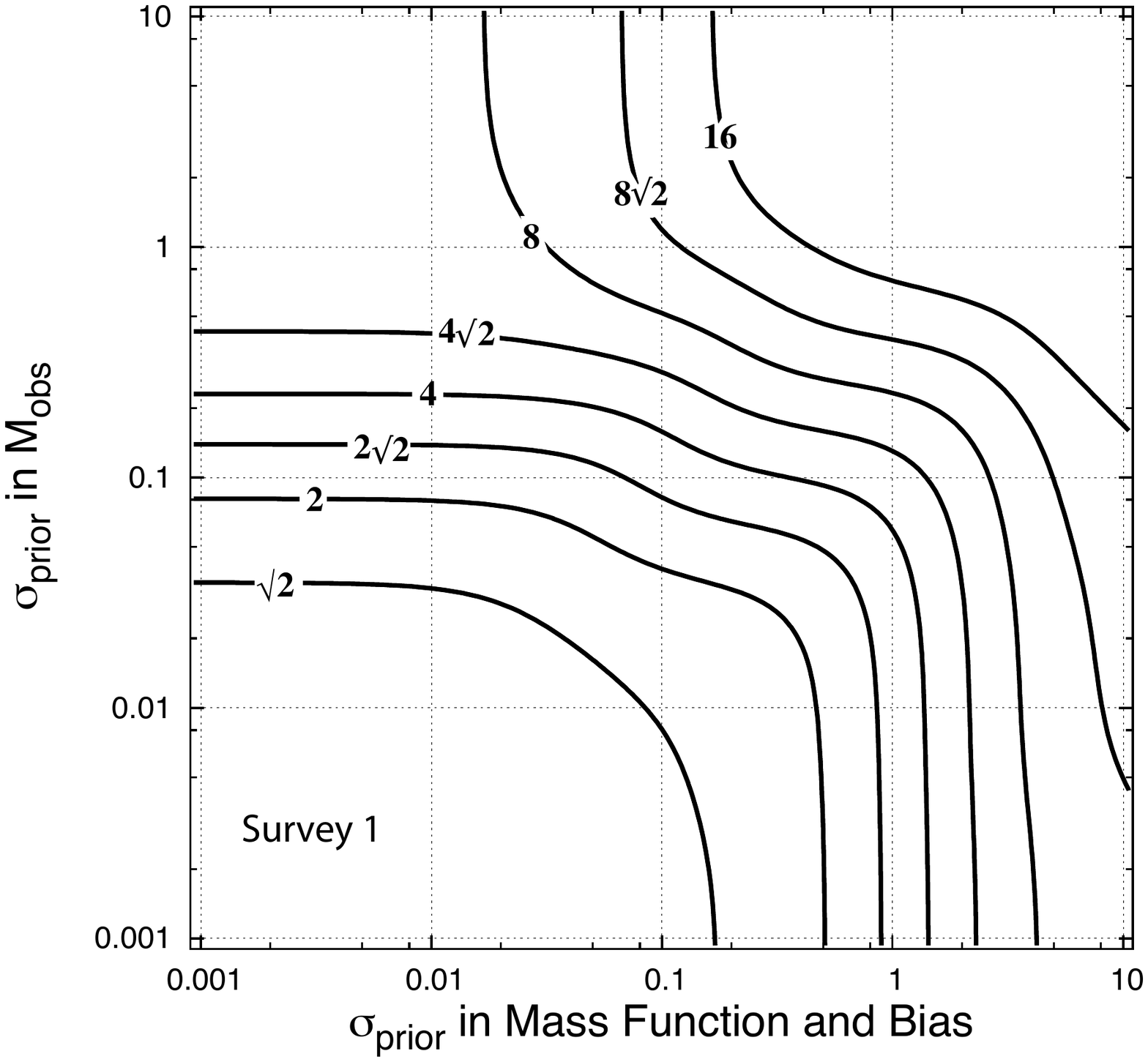}}
    \end{center}
  \end{minipage}
  \begin{minipage}[t]{85mm}
    \begin{center}
      \resizebox{100mm}{!}{\includegraphics[angle=0]{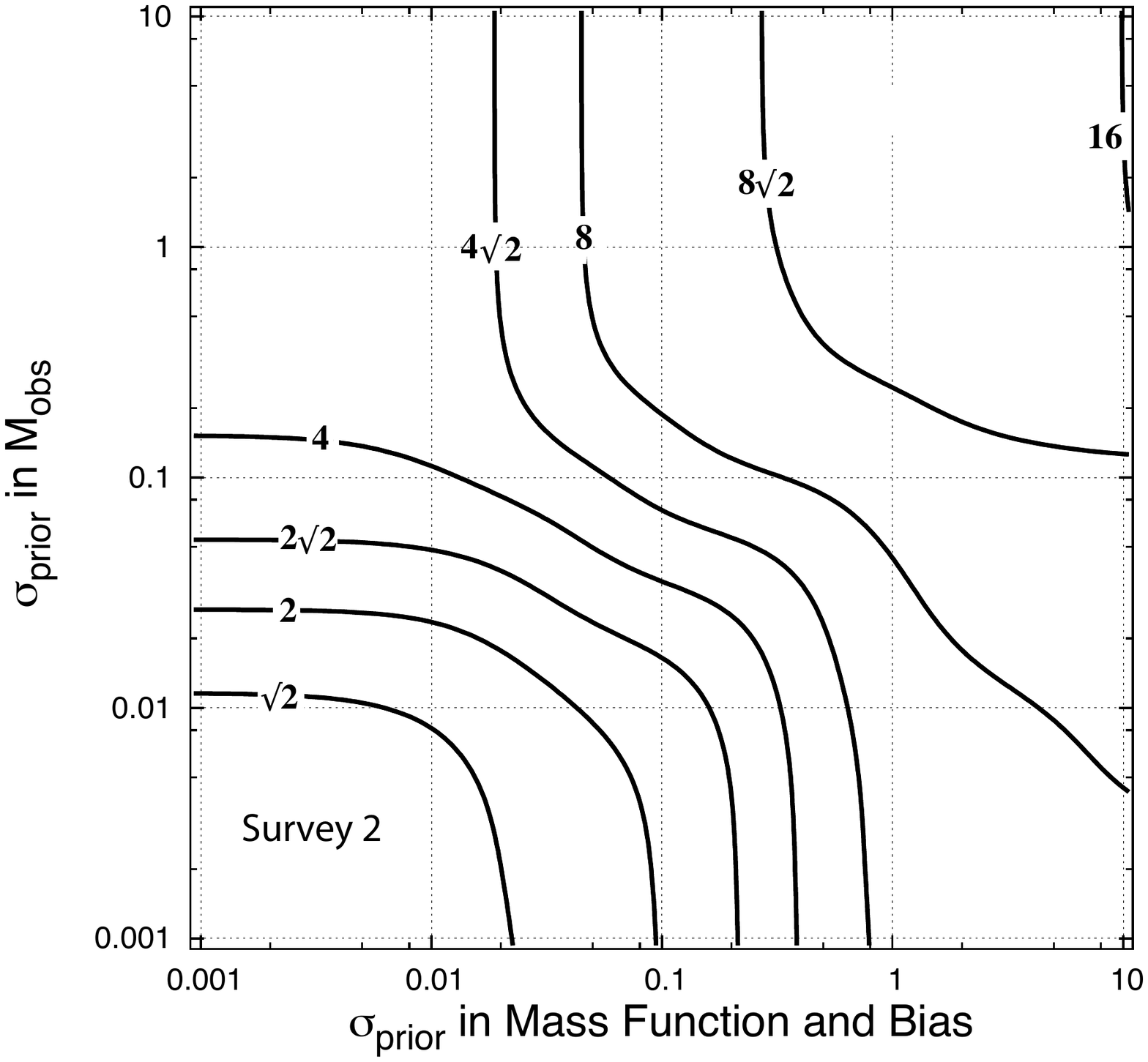}}
    \end{center}
  \end{minipage}
  \begin{minipage}[t]{85mm}
    \begin{center}
      \resizebox{100mm}{!}{\includegraphics[angle=0]{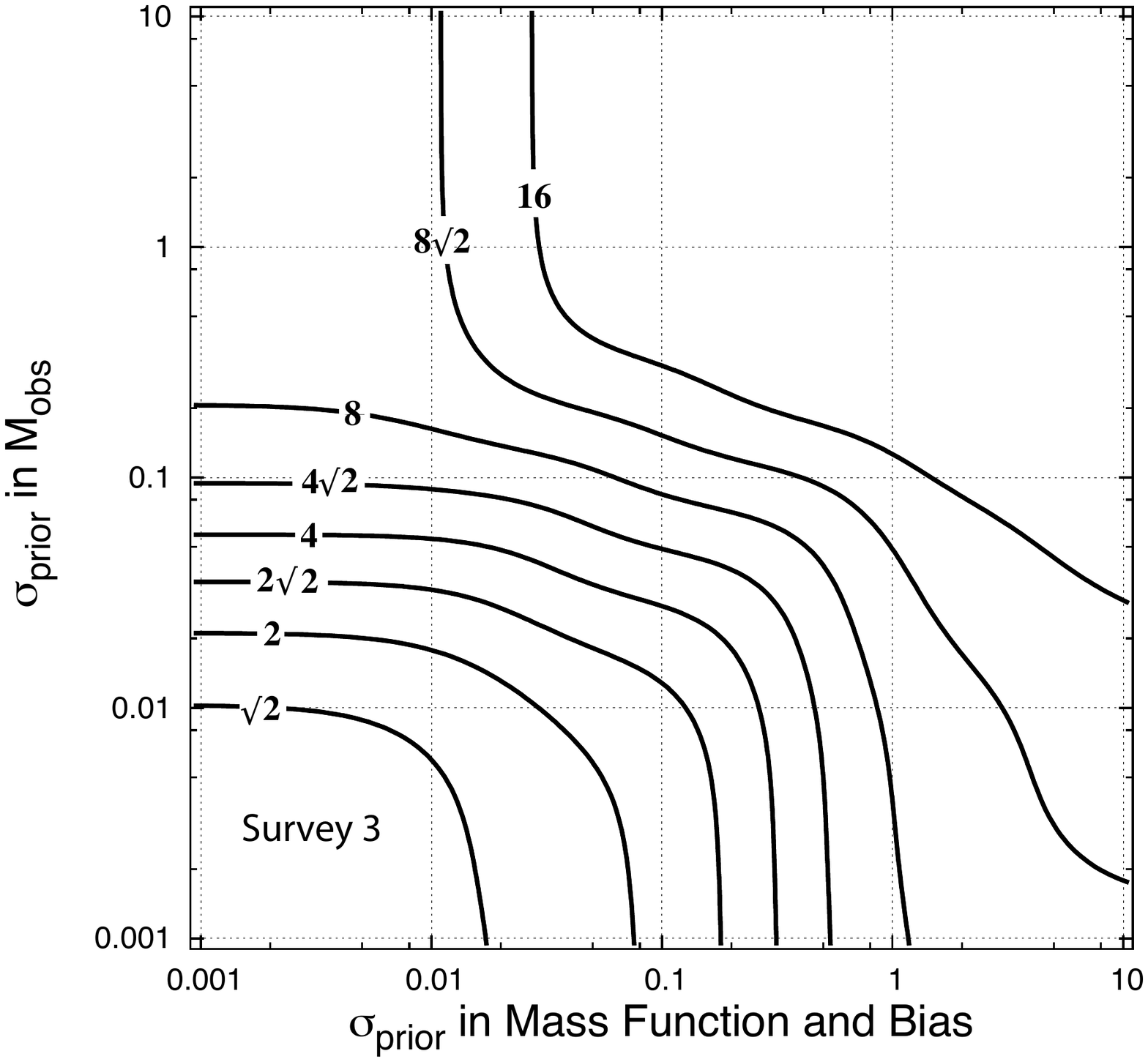}}
    \end{center}
  \end{minipage}
  \caption{Plots of fractional degradation in $\sigma(\DE)$  for
    the Fiducial Survey ({\it top left}), Survey 1 ({\it top right}), Survey 2 ({\it bottom left}), and Survey 3 ({\it bottom right}).
    The contours correspond to regions where constraints on $\Omega_{DE}$
    are degraded by factors of $\sqrt{2}$, $2$, $2\sqrt{2}$, $4$, $4\sqrt{2}$, $8$, $8\sqrt{2}$, $16$, 
    relative to the case of perfectly known nuisance parameters. See Table \ref{tbl:constraints} for the
    baseline constraints on $\DE$.
  }\label{fig:contour.de}
\end{figure*}

\begin{figure*}
    \begin{minipage}[t]{85mm}
    \begin{center}
      \resizebox{100mm}{!}{\includegraphics[angle=0]{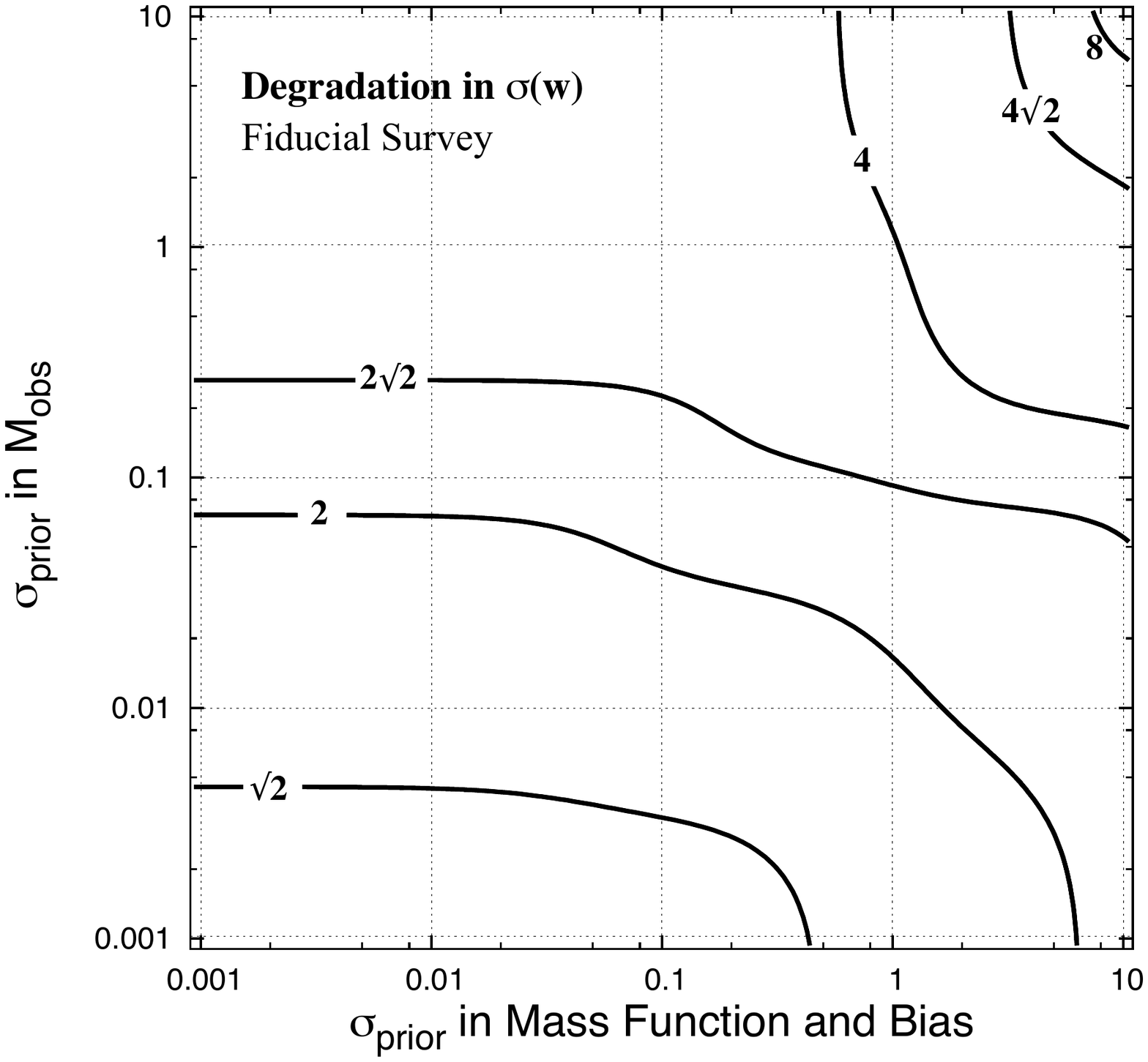}}
    \end{center}
  \end{minipage}
  \begin{minipage}[t]{85mm}
    \begin{center}
      \resizebox{100mm}{!}{\includegraphics[angle=0]{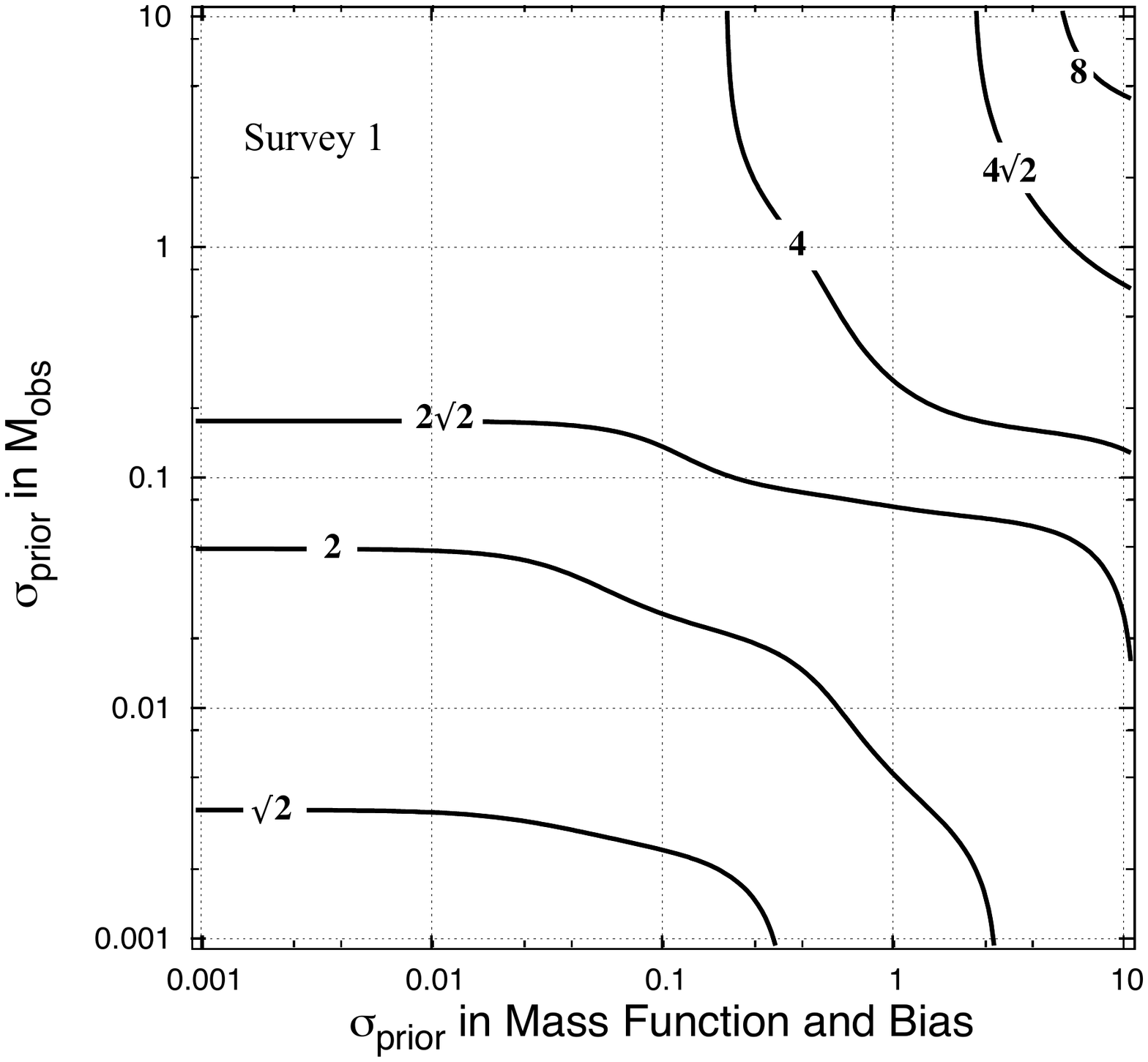}}
    \end{center}
  \end{minipage}
  \begin{minipage}[t]{85mm}
    \begin{center}
      \resizebox{100mm}{!}{\includegraphics[angle=0]{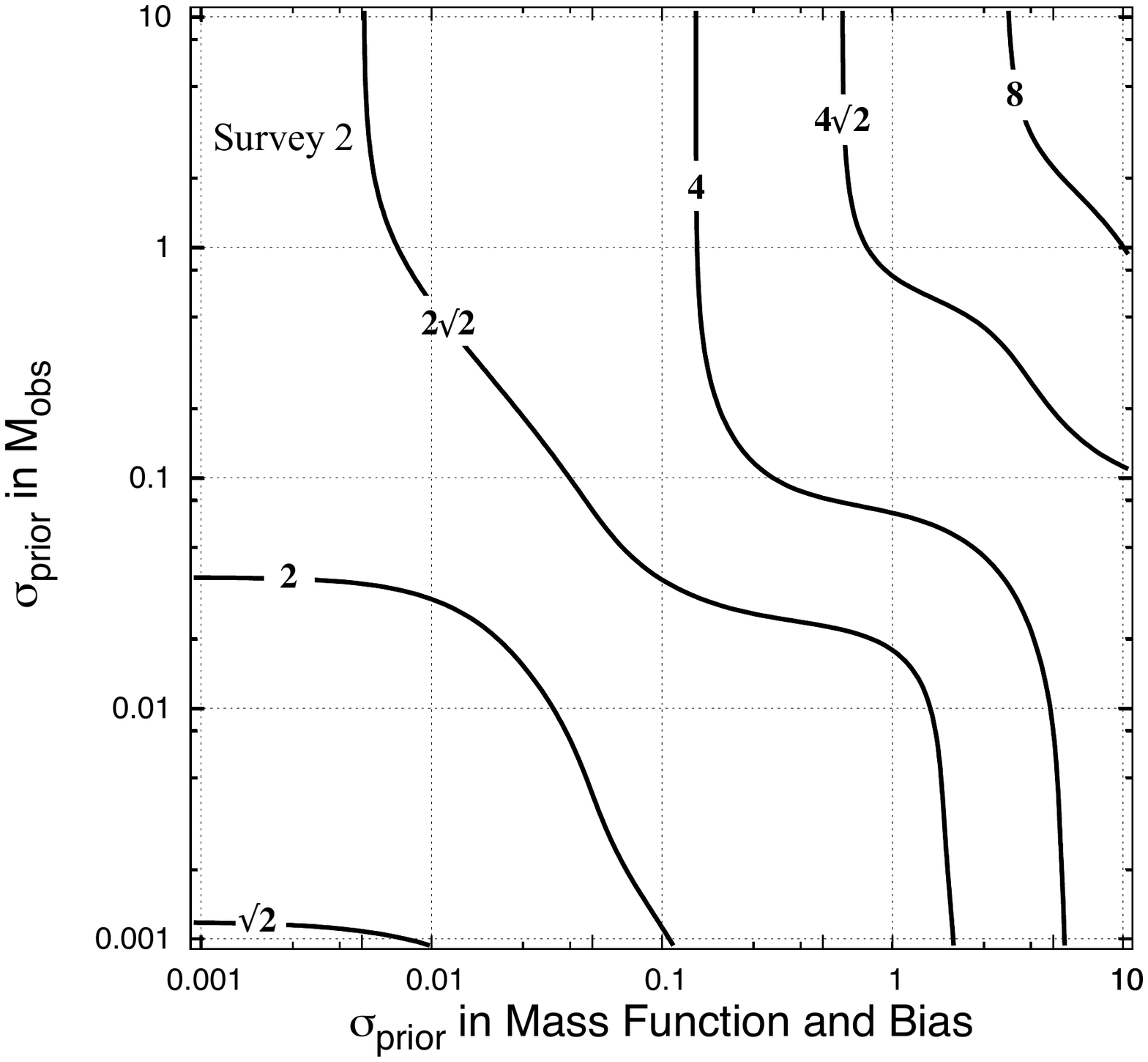}}
    \end{center}
  \end{minipage}
  \begin{minipage}[t]{85mm}
    \begin{center}
      \resizebox{100mm}{!}{\includegraphics[angle=0]{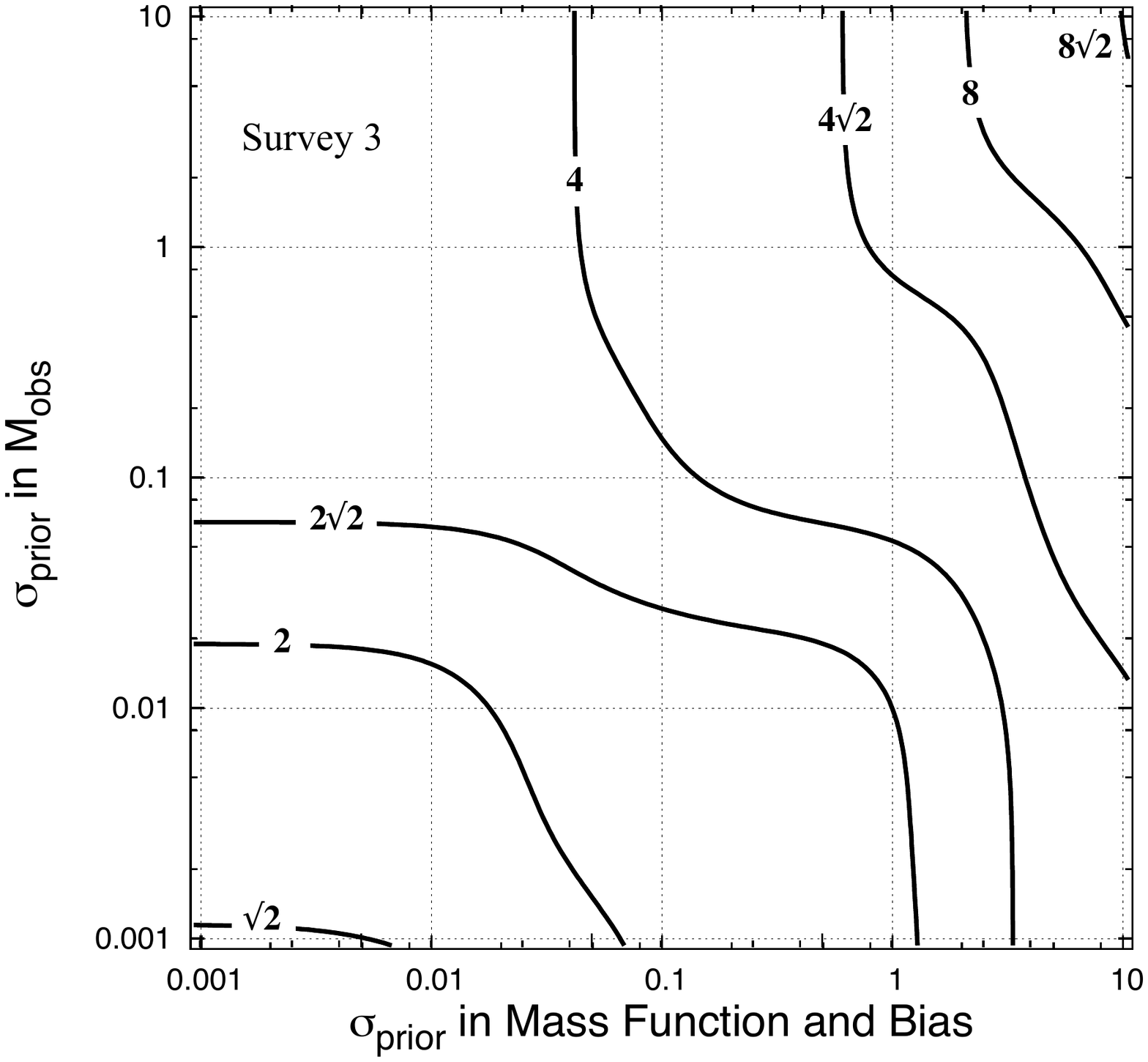}}
    \end{center}
  \end{minipage}
   \caption{Plots of fractional degradation in $\sigma(w)$  for
     the Fiducial Survey ({\it top left}), Survey 1 ({\it top right}), Survey 2 ({\it bottom left}), and Survey 3 
     ({\it bottom right}).
     The contours correspond to regions where constraints on $w$ are degraded by 
     factors of $\sqrt{2}$, $2$, $2\sqrt{2}$, $4$, $4\sqrt{2}$, $8$, $8\sqrt{2}$, $16$, 
     relative to the case of perfectly known nuisance parameters. 
     See Table \ref{tbl:constraints} for the baseline constraints on $w$.
   }\label{fig:contour.w}
 \end{figure*}

Figs. \ref{fig:contour.de} and \ref{fig:contour.w} show contours of the 
multiplicative increase in the errors $\sigma(\DE)$ and $\sigma(w)$, 
respectively, relative to the baseline ``Sharp prior'' constraints given in 
Table \ref{tbl:constraints}.   
The top left panel shows the fiducial survey, with survey 1 (upper right), 
2 (lower left) and 3 (lower right) also shown.  
In all panels, contours show degradation of the error in $\DE$ or $w$ by 
factors of $2^{j/2}$, with $j$ running from 1 to 8.

The contours in Figs. \ref{fig:contour.de} and \ref{fig:contour.w} 
display similar shapes.   Contours tend to intersect the axes at right angles because of the very 
strong prior ($10^{-3}$) being imposed on one sub-space of parameters.   
The contour spacing gives the degradation in constraints as a function of prior 
on the complementary sub-space.   
For small systematic errors, the constraints on $\DE$ and, especially, 
$w$ degrade faster in the $\Mobs$ direction than in the MF/B direction.  
Since the $\Mobs$ systematic degrees of freedom only have redshift, 
not mass, dependence, this indicates a larger sensitivity to redshift 
evolution, particularly in the case of $w$.  
For example, consider a $\sqrt{2}$ increase in $\sigma(w)$ for the 
default survey (upper-left panel of  Fig.~\ref{fig:contour.w}).  
For the case of perfect knowledge of the mass--observable priors, 
this degradation is reached when the mass function uncertainties 
are at the fractional level of $\sim 50\%$.   
Contrast this to the case of perfect knowledge of the mass function 
priors, for which a $\sqrt{2}$ increase in $\sigma(w)$  occurs with 
only $0.5 \%$ error in the mass--observable parameters.  

For the case of $\DE$ shown in Fig. \ref{fig:contour.de}, the sensitivity 
to priors in $\Mobs$ and MF/B is more balanced for small values 
of $\sigo$ and $\sigf$.  
For the fiducial survey, errors of $\sim 0.1$ in the combined model 
parameters produce a factor 2 increase in $\sigma(\DE)$.  
For $\sigo \gtrsim 1.0$, the sensitivity of the degradation 
to priors on $\Mobs$ parameters decreases sharply.  
This plateau reflects the ability of mass function shape and clustering information alone 
to jointly calibrate $\Mobs$ parameters and $\DE$ constraints.
The same effect is noticeable in Fig. \ref{fig:contour.w}, though for $w$, the
shape information provides the primary source of constraints of 
the $\Mobs$ parameters. 
When $\sigf$ is below the level of a few percent, the survey self-calibration
is more effective at constraining the $\Mobs$ nuisance parameters, so that
increasing $\sigo$  above $\sim 0.1$ does not result in 
further degradation of the cosmological constraints.
The corresponding plateau in the MF/B parameter direction is much less 
pronounced. 
Consequently, if no prior information is available, mass-function/bias 
uncertainties dominate the error budget in cosmological parameters.

The contours of fixed degradation shift as one considers the other surveys 
shown in the remaining panels of Figs. \ref{fig:contour.de} and \ref{fig:contour.w}.   
Comparing the right panels (showing surveys with $\sigma_0 = 0.5$), to the left (surveys with $\sigma_0=0.2$), we 
see that increasing the default $\Mobs$ scatter tends to shift the contours inwards, signifying an increase in 
sensitivity to the priors.   
This increase in sensitivity offsets the smaller baseline error in 
$\DE$ and $w$, leading to roughly constant errors in these parameters for fixed nuisance priors.
The shift inwards is most noticeable for large values of $\sigp$ for the surveys with 
$\Mth=10^{13.5} h^{-1}\Msun$.
In the limit of flat priors on both MF/B and $\Mobs$ parameters, we see from
Table \ref{tbl:constraints} that the increase in scatter is beneficial
for the surveys with $\Mth=10^{14.2} h^{-1}\Msun$ but detrimental to 
the surveys with $\Mth=10^{13.5} h^{-1}\Msun$.
The former are dominated by shot noise, hence benefit from the increase in 
counts, but the latter are dominated by sample variance.

Comparing the bottom panels (surveys with $\Mth=10^{13.5} h^{-1}\Msun$) with those above (surveys with 
$\Mth=10^{14.2} h^{-1}\Msun$), we see that 
the overall effect of decreasing the 
mass threshold is to increase the sensitivity to MF/B and $\Mobs$ priors.    
Both $\DE$ and w degradation contours shift inwards by as much as an order of magnitude.  
The effect is most pronounced in the MF/B direction.
For the case of $w$ constraints with sharp $\Mobs$ priors, an intermediate 
plateau emerges in the range $\sigf \sim 0.1 -1$. 
With lower $\Mth$, there is more information in cluster surveys, which requires 
very accurate priors to fully extract.

In summary, surveys with lower (better) baseline dark energy constraints have 
tighter  requirements for priors on model systematic effects.
For small $\sigo$ and $\sigf$ ($\lesssim 1$), the larger degradation for tighter
 baselines nearly offset,  in the sense that the absolute uncertainties in $\DE$ 
and $w$ at fixed $\sigo$ and $\sigf$ are almost constant among the surveys.
For  high $\sigp$ the better baseline does not compensate the more stringent prior 
requirement because the surveys with larger scatter are more sensitive to the prior uncertainties.

These results assume no mass-dependent evolution of the mass--observable relation. 
If we adopt the parameterization of \cite{cun08} which consists of adding one parameter
to describe the mass-evolution of the bias and three parameters to characterize the mass-evolution
of the scatter, we do not find significant qualitative changes.
In survey 3, for example, assuming $\sigf=\sigo=0.1$ the four additional parameters yield
essentially no shift in the contours. 
In other regions of the Figure, results can be more noticeable, with shifts in the $\sigf$ or $\sigo$
directions of up to a factor of 2. 
The rough shape of the curves is very similar.
Degradations of $\sigma(w)$ are somewhat more sensitive to the inclusion of mass-dependent
nuisance parameters. 
Assuming $\sigf=\sigo=0.1$ in Survey 3, addition of mass-dependence degrades constraints by roughly $\sqrt{2}$.
The effects on the other surveys are even weaker since either they probe a smaller range in mass or they
have smaller fiducial scatter.

\begin{figure*}
      \resizebox{85mm}{!}{\includegraphics[angle=0]{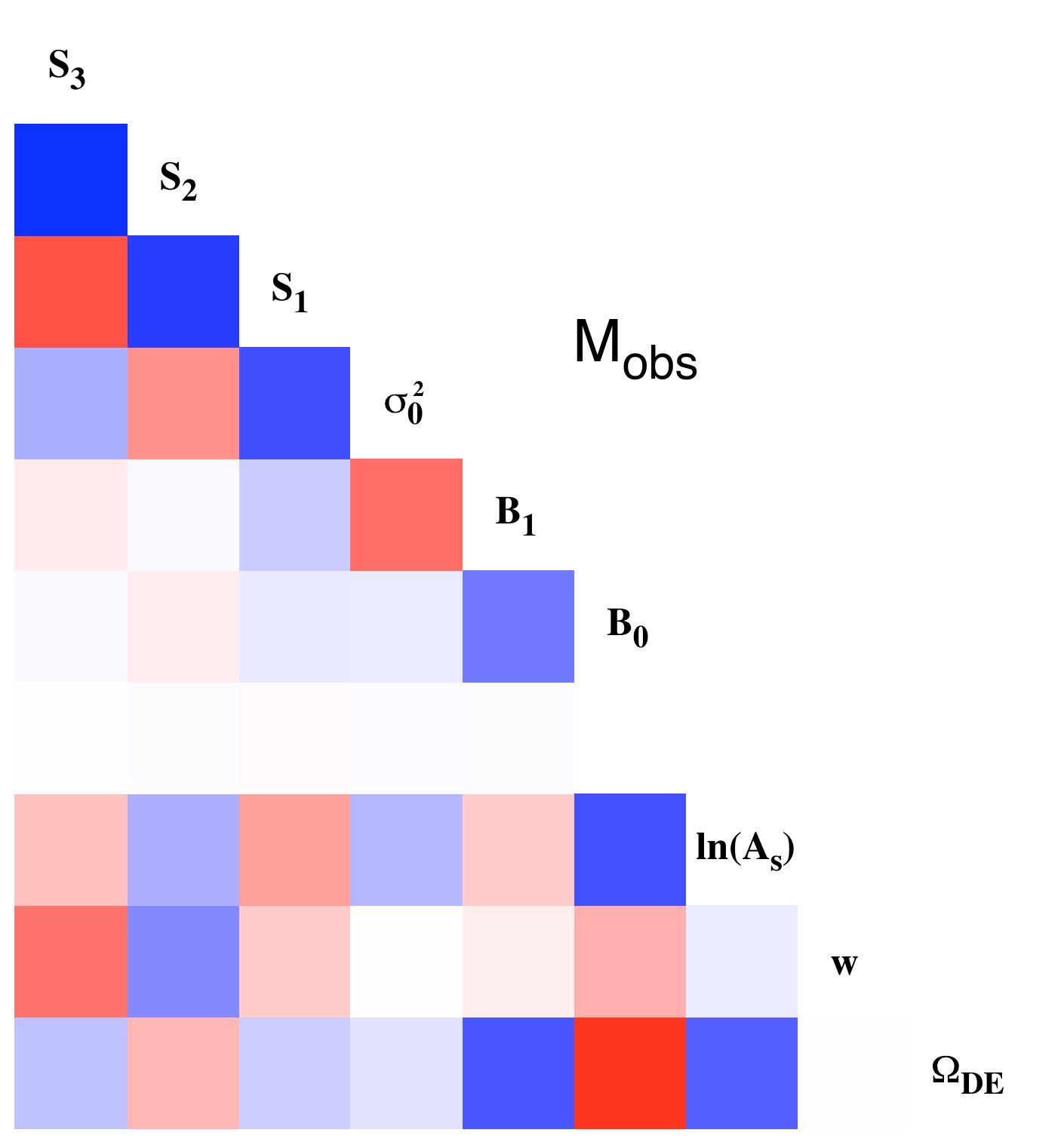}}
      \resizebox{85mm}{!}{\includegraphics[angle=0]{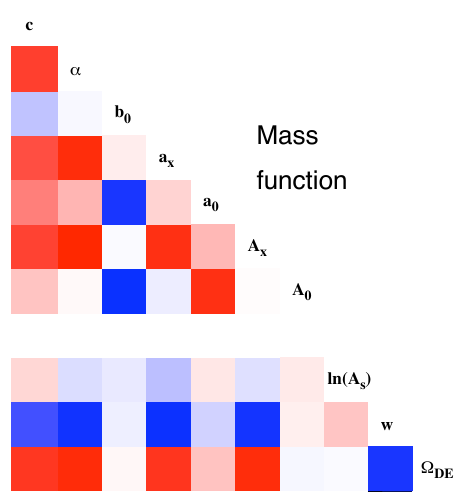}}
      \resizebox{155mm}{!}{\includegraphics[angle=0]{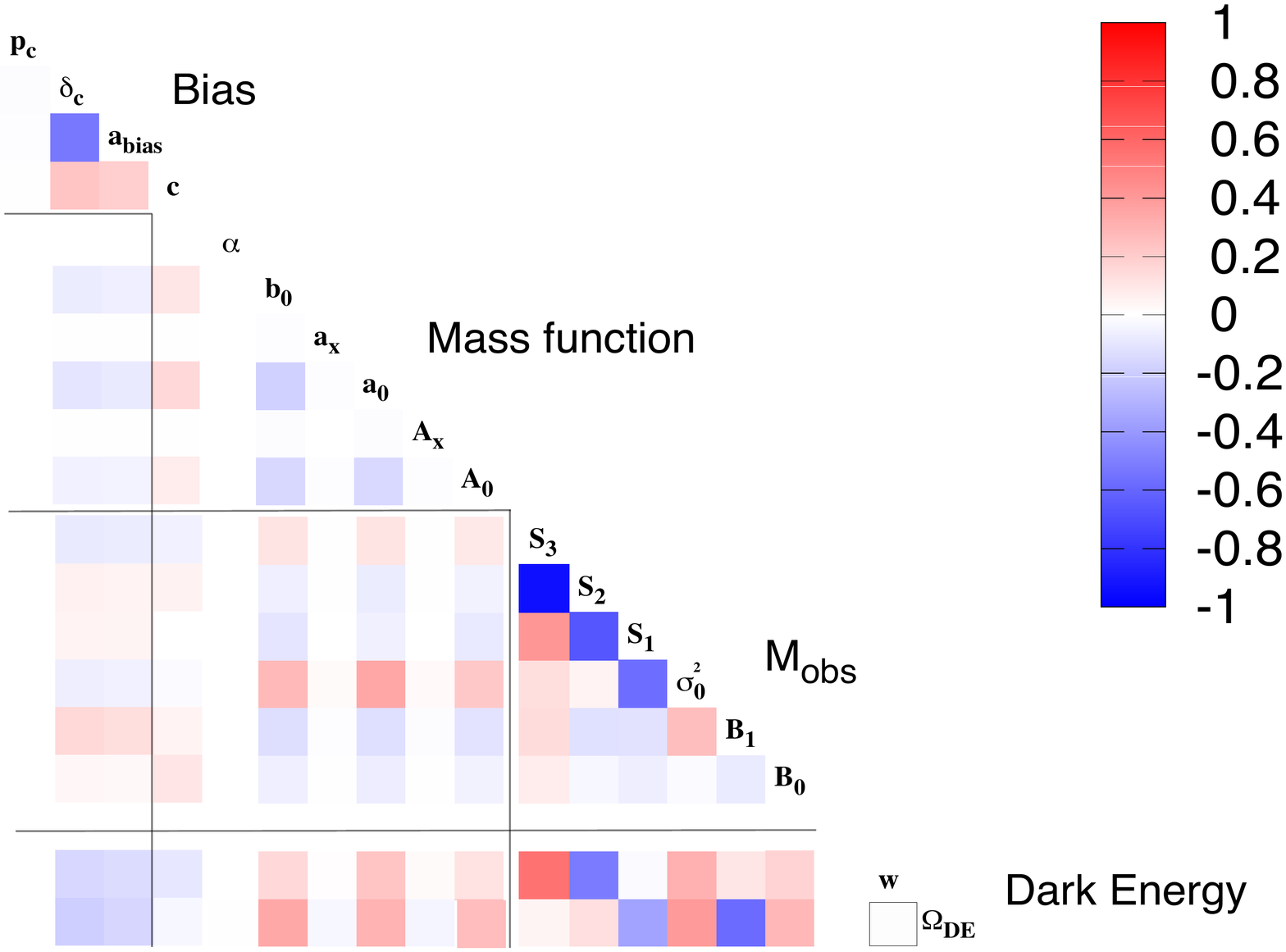}}
  \caption{ Correlations between cosmological parameters and ({\it top left}) 
$\Mobs$ nuisance parameters, and ({\it top right}) MF/B nuisance parameters. 
The nuisance parameters not shown in each plot are fixed by infinitely sharp priors. 
Cosmological parameters, with the exception of $\DE$ and w have $1\%$ priors and 
are not shown unless significant correlation with other parameters is present. 
The {\it bottom} plot shows the correlation between $\Mobs$, MF/B, $\DE$, and w. 
We have added priors to all nuisance parameters such that $\sigo=\sigf=0.1$, which 
very roughly corresponds to our present knowledge of these parameters. 
  }\label{fig:correlation}
  \end{figure*}

\subsection{Parameter correlations}\label{sec:correl} 

The halo modeling nuisance parameters have complex correlations among themselves 
and with the cosmological parameters.  
While one could imagine adopting a more orthogonal parameterization with potentially fewer parameters, 
the interpretation of the results would be hard to relate to the functional forms
currently in use.
Thus, we pursue the direct approach of exploring the Fisher correlations of the full parameter set.   
We begin by isolating the $\Mobs$ and the MF/B sub-spaces separately, using 
sharp priors on the complementary sub-space, then consider the full parameter covariance 
matrix under the assumption of a 10\% prior uncertainty on the halo modeling parameters.

First, we add sharp priors to the MF/B nuisance parameters and consider the correlations 
between the $\Mobs$ nuisance parameters and the dark energy parameters, 
shown in Fig.~\ref{fig:correlation} ({\it top left}). 
The labels of the $\Mobs$ and cosmological parameters occupy the diagonal of the 
correlation matrix.   
As described in Sec. \ref{sec:clustermass} the cosmological parameters other than $\DE$ 
and $w$ have $1\%$ priors, and this level is small enough to remove almost all of their 
correlations with the other parameters.  
The exception is the normalization of the primordial power spectrum, $\ln(A_s)$, which strongly 
correlates with $\DE$ despite the $1\%$ prior.

From the plot, we see that the constant nuisance parameter of the mass bias ($B_0$) is strongly 
correlated with $\DE$ whereas the redshift-related parameter ($B_1$) is markedly anti-correlated 
with $\DE$.  The variance nuisance parameters ($\sigma_0^2$, $S_1$, $S_2$, $S_3$) do not 
correlate as strongly with $\DE$.  
In contrast, the equation of state $w$ is mostly correlated with the redshift-dependent 
variance parameters, and somewhat correlated to $B_0$.  
These correlations are consistent with the trends seen in Fig. 15 of \cite{cun08}, 
showing the dependence of $\DE$ and w constraints on the priors on different $\Mobs$ parameters.

On the ({\it top right}) plot of Fig. \ref{fig:correlation} we examine the correlations between 
the mass function nuisance parameters and the cosmological parameters.
As in the above case, we apply infinitely sharp priors to all nuisance parameters not shown, and only plot
the parameters with significant correlations.
In this scenario, the $\ln(A_s)$ correlates more weakly with all other parameters.
The redshift-dependent parameters ($\alpha$, $a_x$, $A_x$) show strong positive correlations with $\DE$,
and strong negative correlations with $w$.
The exponential cutoff parameter $c$ is also noticeably correlated with $w$ and $\DE$.
The redshift-dependent nuisance parameters are also strongly correlated among themselves, as are the
constant parameters ($A_0$, $a_0$, and $b_0$), which are very weakly correlated with all other parameters.  

In the bottom panel of Fig. \ref{fig:correlation}, we investigate the full correlation matrix after imposing 
priors corresponding to  $\sigo=\sigf=0.1$ to the $\Mobs$ and MF/B parameters (along with 
the $1\%$ cosmological priors). 
The $\ln(A_s)$ correlations are very weak, hence, we do not show them.  
The choice of priors on the halo modeling parameters is admittedly crude, based roughly on the present-day 
understanding of the mass-function, bias, and the mass--observable relation.   
We caution that the correlations are a strong function of the imposed priors, so the chosen case is 
illustrative rather than definitive.  

With these priors, the uncertainties in the $\Mobs$ nuisance parameters dominate 
the error budget in $\DE$ and $w$.   This result is consistent with the shape of the contours near 
$(0.1, \, 0.1)$ in Figs. \ref{fig:contour.de} and \ref{fig:contour.w}.   
Thus, the correlations between the $\Mobs$ and dark energy parameters are more pronounced,
and, in general, resemble the correlations, performed under sharp MF/B priors, 
displayed in the top left panel.  
An exception is the correlation between the mass bias constant, $B_0$, and $\DE$, 
for which the correlation in the full treatment is substantially weaker.  

The mass function parameter correlations are substantially different from their isolated treatment.   
In particular, the redshift evolution parameters ($A_x$, $a_x$, and $\alpha$) largely disappear, 
and only the constant parameters ($A_0$, $a_0$ and $b_0$) contribute appreciably 
to the dark energy error budget.   
That this behavior differs from the isolated case is not too surprising.  
In the isolated case, the $\Mobs$ parameters are assumed to be perfectly known, so 
the only redshift evolution remaining in the model to compete with dark energy is 
contained in the MF/B parameters.   
In the full case, the assumed prior level of $\Mobs$ uncertainty is sufficient to dominate 
the evolutionary behavior, leaving the primary shape parameters of the MF as the means by 
which this sector affects dark energy constraints.  

The full analysis also includes halo bias nuisance parameters.  
The parameters $\delta_c$ and $a_{\rm bias}$ strongly correlate with each other and 
also display weakly negative correlations with both $\DE$ and $w$.  
The parameter $p_c$ is virtually uncorrelated with all the other parameters.

As we have mentioned before, the detailed behaviors are a consequence of the priors applied to the 
nuisance parameters.  Had we used flat (very weak priors), the MF/B nuisance parameters would dominate the 
error budget, since self-calibration provides some constraints on the $\Mobs$ parameters.
Then, the correlations of the mass function parameters between themselves 
and the cosmological parameters would more closely resemble those seen at the top right panel.
The bias parameters would also exhibit somewhat stronger correlations.

\begin{figure*}
   \begin{minipage}[t]{85mm}
    \begin{center}
      \resizebox{85mm}{!}{\includegraphics[angle=0]{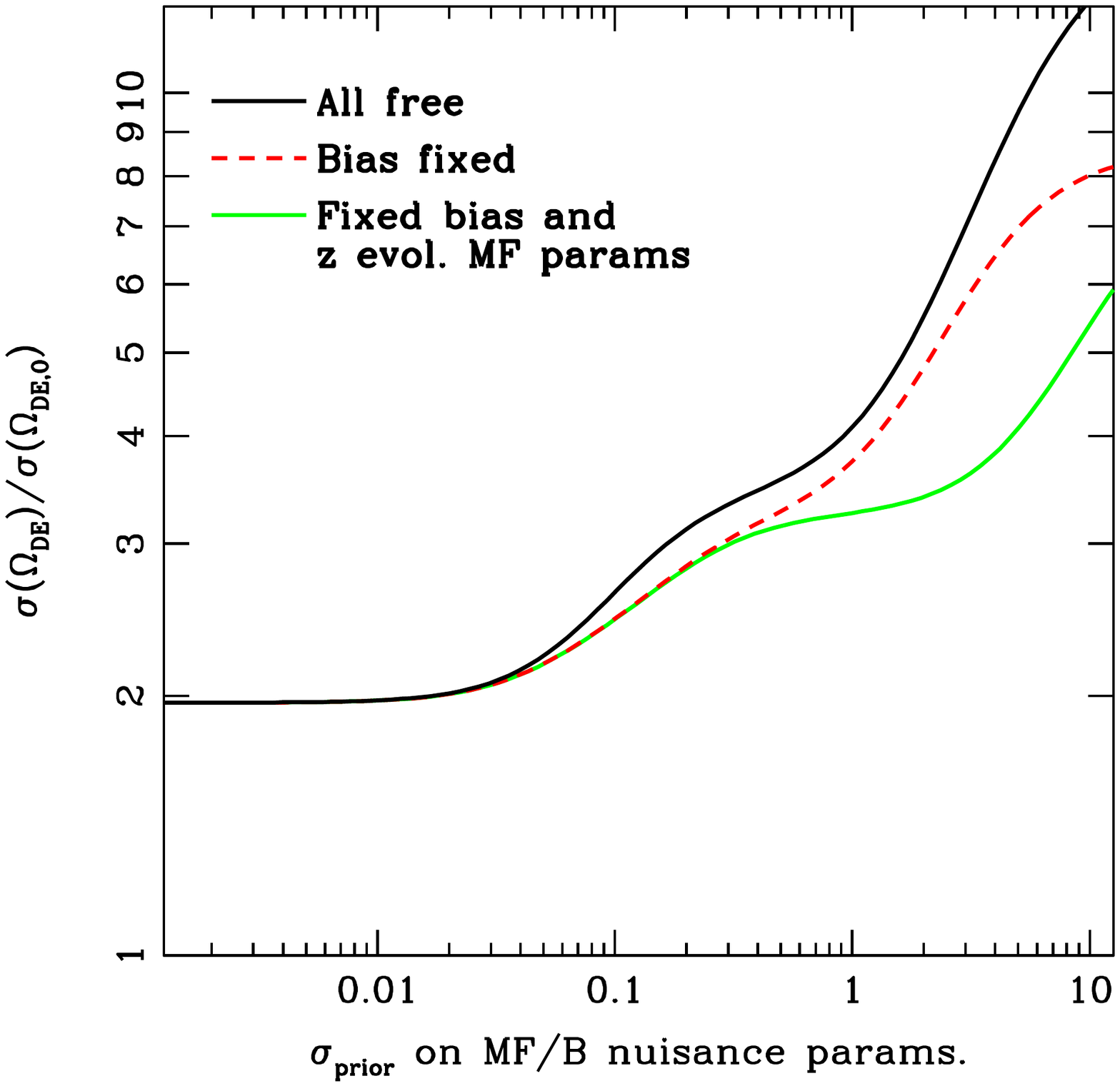}}
    \end{center}
  \end{minipage}
 \vspace{-2.5cm}
  \begin{minipage}[t]{85mm}
    \begin{center}
      \resizebox{85mm}{!}{\includegraphics[angle=0]{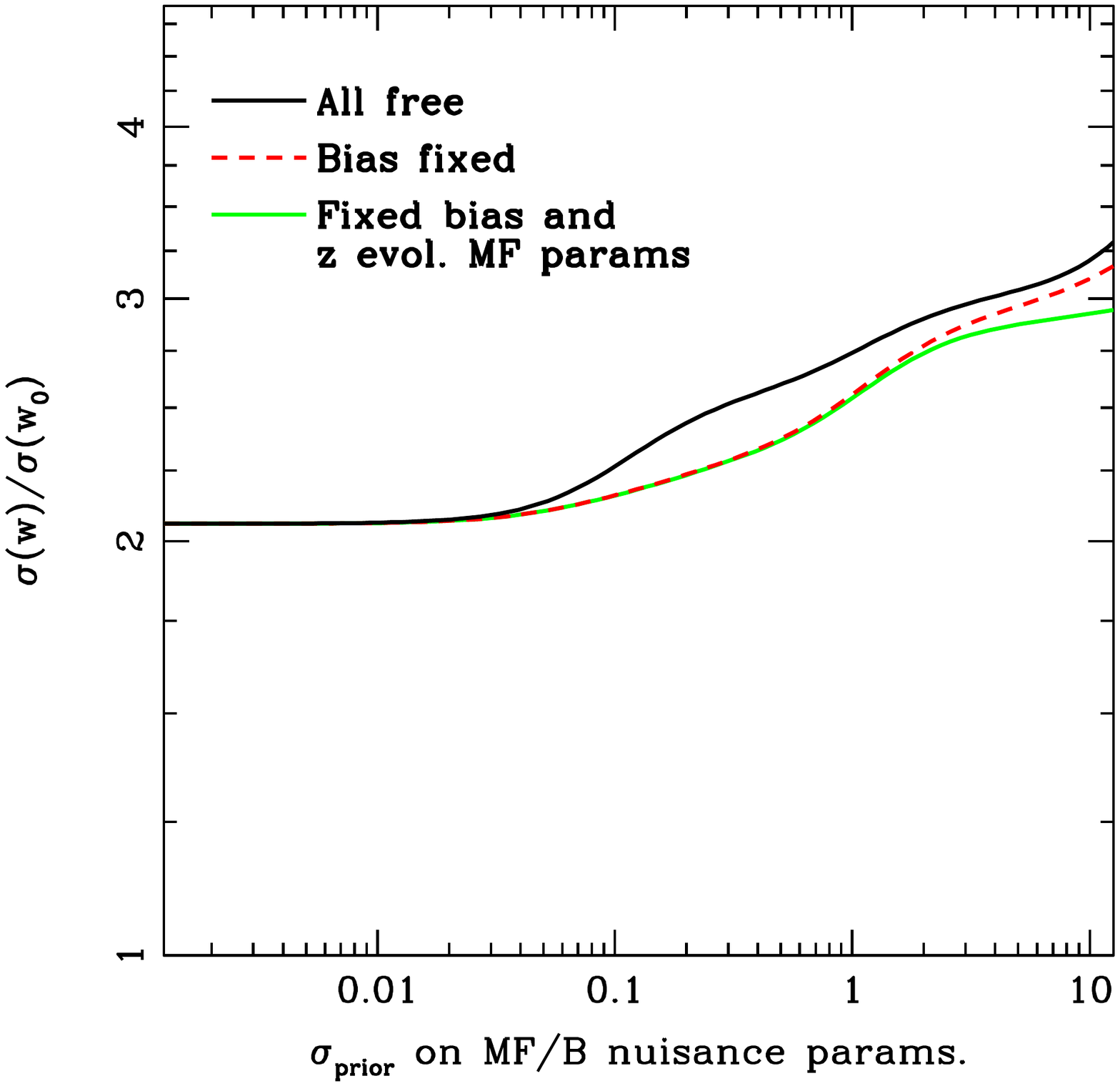}}
    \end{center}
  \end{minipage}
 \vspace{-2.5cm}
   \begin{minipage}[t]{85mm}
    \begin{center}
      \resizebox{85mm}{!}{\includegraphics[angle=0]{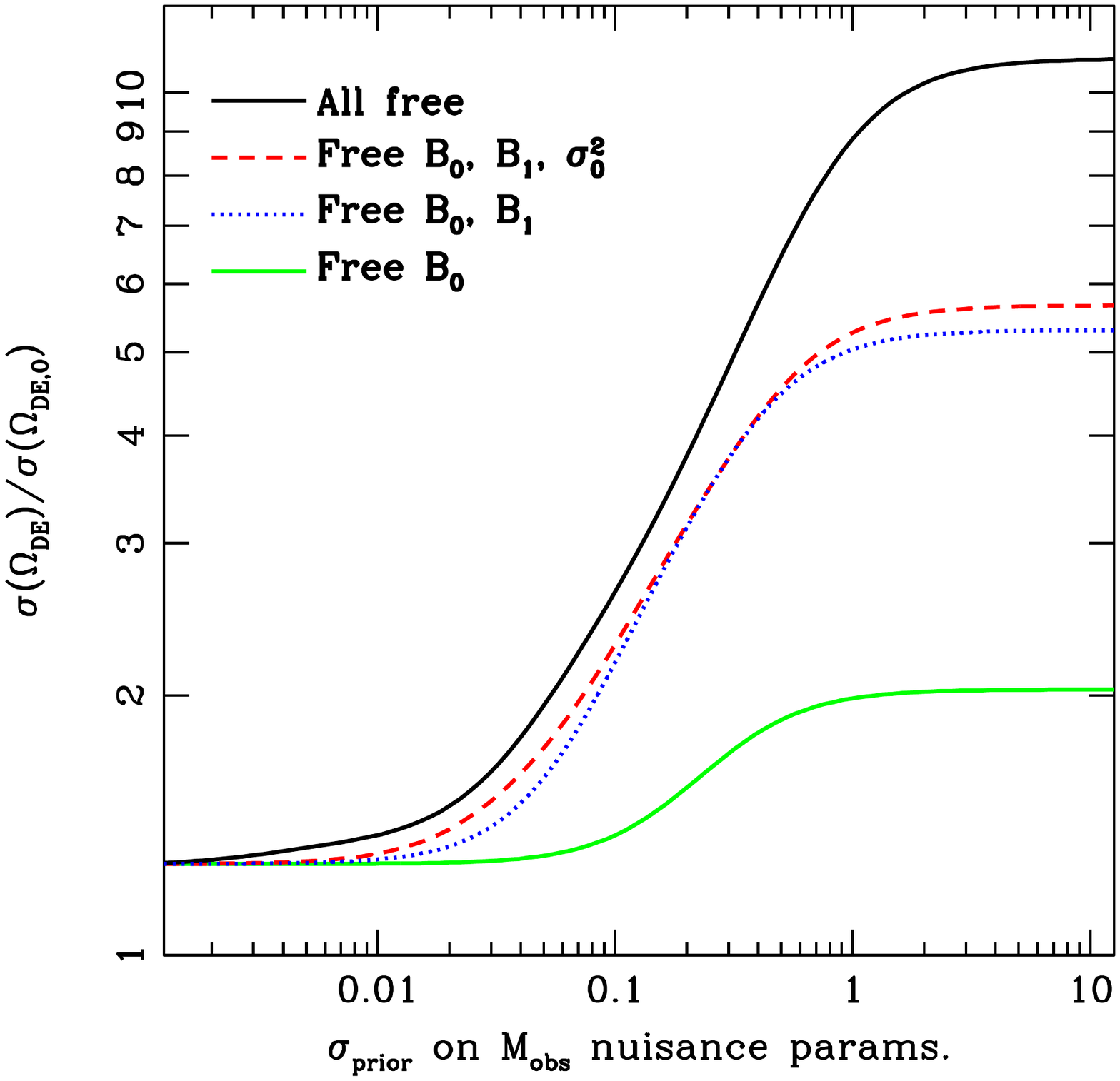}}
    \end{center}
  \end{minipage}
  \begin{minipage}[t]{85mm}
    \begin{center}
      \resizebox{85mm}{!}{\includegraphics[angle=0]{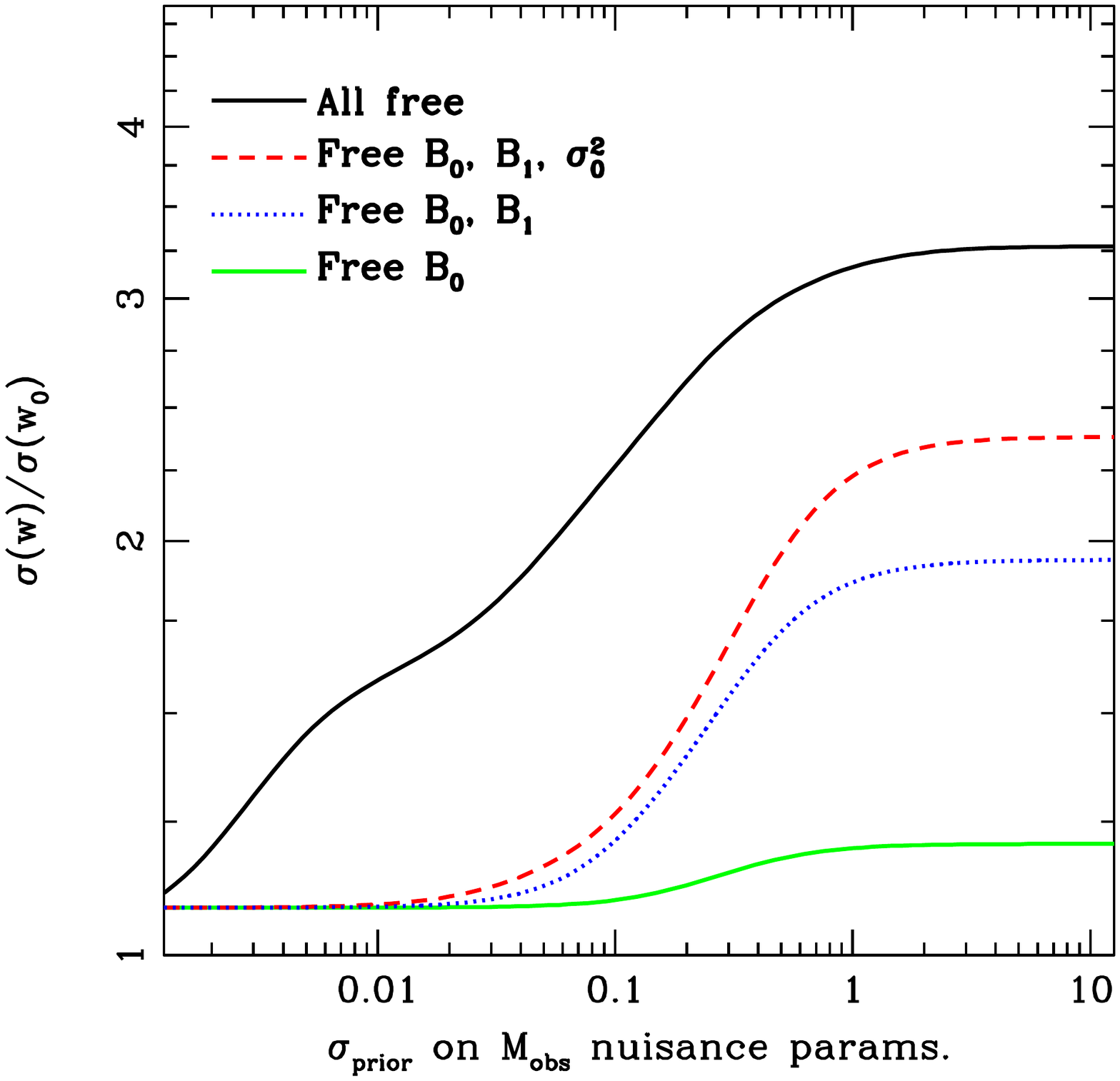}}
    \end{center}
  \end{minipage}
  \vspace{1.0cm}
  \caption{Degradation of constraints on ({\it left}) $\Omega_{DE}$ 
    and ({\it right}) equation of state $w$ as a function of the prior on the
    uncertainty in MF/B and $\Mobs$ nuisance parameters: $\sigf$ and $\sigo$, respectively 
(defined in Eqs. \ref{eqn:primobs} and \ref{eqn:primf}).
    Plots in the {\it top row} assume prior uncertainty on the $\Mobs$ parameters of $\sigo=0.1$.
    The {\it solid black} lines assume no priors on any of the MF/B nuisance parameters.
    The {\it dashed red} lines assume sharp priors on the three bias parameters, and 
    the {\it solid green} lines assume sharp priors on the redshift evolution mass function
    nuisance parameters ($A_x$, $a_x$, and $\alpha$).
    In the range of $\sigf$ plotted, the {\it green} line is unaffected if priors are 
    applied to the bias parameters or not. 
    Plots in the {\it bottom row} assume prior uncertainty on the MF/B parameters of $\sigo=0.1$.
    The {\it solid black} lines assume no priors on any of the $\Mobs$ nuisance parameters.
    The {\it dashed red} lines assume sharp priors on the three parameters describing the 
redshift evolution of the mass variance. 
    The {\it dotted blue} lines assume sharp priors on all four mass variance
    nuisance parameters ($\sigma_0^2$, $S_1$, $S_2$, and $S_3$), and the {\it solid green} lines
assume sharp priors on all $\Mobs$ nuisance parameters except the constant bias term $B_0$.
  }\label{fig:degrad}
 \end{figure*}

To further illustrate this sensitivity to priors on nuisance parameters,
we explore the contributions of different sets of MF/B and $\Mobs$ nuisance parameters to degradations
in $\sigma(\DE)$ and $\sigma(w)$. 
The top row of Fig. \ref{fig:degrad} shows the fractional degradation of $\sigma(\DE)$ and $\sigma(w)$ with 
respect to the baseline of the fiducial survey for the cases where (1) all the MF/B parameters
are allowed to vary, (2) only the MF parameters are free, and (3) only the redshift-independent
MF parameters are free. 
As before, we fix the priors on $\Mobs$ parameters so that $\sigo=0.1$.

In the {\it upper left} plot of Fig. \ref{fig:degrad}, we see that 
the redshift evolution MF nuisance parameters ($A_x$, $a_x$, and $\alpha$) dominate 
the degradation of $\sigma(\DE)$ for moderate to large error, $0.1 \lesssim \sigf \lesssim 1$. 
Below $\sigf \sim 0.1$, the constant MF parameters ($A_0$, $a_0$, $b_o$ and $c$) are 
the most relevant.  The bias parameters are relevant in the high-uncertainty range, $1 \lesssim \sigf \lesssim 10$.
When $\sigf=0.1$, the parameters not related to redshift evolution, {\sl i.e.}, $A_0$, $a_0$, and $b_0$ 
dominate, as seen in Fig. \ref{fig:correlation}.
In the {\it upper right} plot, the constant MF nuisance parameters dominate 
the $w$ constraints up to $\sigf \sim 1$.
For larger $\sigf$, the redshift evolution parameters become more important.
The bias parameters are moderately relevant in the range $\sigf \sim 0.03-1$.

The bottom row of Fig. \ref{fig:degrad} shows the degradation of $\DE$ and $w$ constraints 
as we vary priors on $\Mobs$ parameters while keeping $\sigf$ fixed at 0.1.
From the {\it lower left} plot, we see that the parameters related to the redshift evolution 
of the mass--observable relation are the most important for virtually the entire interval we examine.
In particular, the redshift evolution of the mass bias is the most relevant for $\DE$ constraints.
In the {\it lower right} plot we see that the redshift evolution of the mass variance
dominates $w$ constraints for small uncertainties in the priors.   For $\sigo \gtrsim 0.1$, the constant part of the
variance, $\sigma_0^2$, as well as the redshift evolution of the bias, dominate.
Constraints are almost independent of priors on the constant bias term, $B_0$, but this term does affect the power 
spectrum normalization, $\ln(A_s)$.

\section{Discussion}\label{sec:disc}
The results presented in this paper make a variety of assumptions of various 
degrees of relevance, which must be interpreted with caution.
In this section we discuss the generality of some of our assumptions.

{\it Parameterization of the mass--observable relation}
We parameterized the redshift evolution of the variance in $\Mobs$ using a cubic 
polynomial.
Lima $\&$ Hu (2005) \cite{lim05} show that the cubic polynomial is almost as complete
a description as having fully independent scatter in $\sim 20$ redshift bins.
Hence, we feel that our parameterization is conservative with regards to redshift evolution. 
There is no physical motivation for this choice, however, and if simpler parameterizations
describe the data well, then constraints would improve - and the sensitivity to the
uncertainty in the $\Mobs$ parameters would decrease.
As a test, we eliminated the quadratic and cubic terms in the scatter in $\Mobs$.
The contour lines of Figs. \ref{fig:contour.de} and \ref{fig:contour.w} shifted upwards 
by as much as factors of 5 in $\sigma_{\rm prior}$. 

We did not include mass evolution of the mass--observable relation since evidence 
supporting this assumption from observations and simulations is currently weak.
For surveys with high $\Mth$, we have checked that cosmological constraints are virtually
unaffected by adding a cubic evolution of the mass scatter plus a linear evolution
of the mass bias. 
For surveys with low $\Mth$, the mass terms cause an increase in sensitivity to both
$\Mobs$ and MF/B parameters, which results in a shifting of the contours of 
Figs. \ref{fig:contour.de} and \ref{fig:contour.w} inwards.
The additional $\Mobs$ nuisance parameters correlate strongly with the MF/B nuisance 
parameters, so that the sensitivity of both sets of parameters to the priors vary
in similar fashion.  

Our results are based on a quite generic procedure for adding priors. 
We did not strive for optimal dark energy constraints, or to accurately reproduce
what specific observations and simulations might return.   
Understanding the optimal priors needed - for fixed observational/simulational costs - can
be very valuable \cite{wu09}, and such detailed studies are important complements to 
the generic treatment presented here. 

As mentioned previously, we did not consider variations in w in this paper. 
The sensitivity of the Dark Energy constraints to the redshift-dependent nuisance parameters
should increase, though it is not obvious that the relative importance of the $\Mobs$ and MF/B
nuisance parameters would change significantly.
The subject of cluster constraints on time-varying w, parameterized as $w_0$/$w_a$, is addressed in 
papers such as Wu et al. \cite{wu09b} and Cunha et al \cite{cun09}. 
 
\section{Conclusions}\label{sec:conc}

We investigate the sensitivity of dark energy constraints from clusters of galaxies
to halo modeling uncertainties in the mass function, clustering bias, and the mass--observable relation. 
We find that mass--observable uncertainties dominate the error
budget for both $\DE$ and $w$ constraints for surveys with higher mass-thresholds,
such as SZ surveys, assuming prior uncertainties of order $0.1$ on both mass--observable
and mass-function/bias (MF/B) nuisance parameters.
For surveys with lower mass-thresholds, the uncertainties in the mass--observable and 
MF/B nuisance parameters are more comparable, depending on the degree of prior knowledge 
assumed.

The variations in the sensitivity to the prior uncertainties are offset by 
the different baseline constraints of each survey.  
Not surprisingly, surveys with lower (better) baseline constraints have 
tighter requirements for priors on model systematic effects.

We examine the correlations between the nuisance and cosmological parameters for the
fiducial survey for several different assumptions about prior knowledge of the nuisance
parameters. 
If the mass--observable relation is perfectly known, the mass function parameters show
strong positive (negative) correlations with $\DE$ and w.
If the mass function parameters are known to $\sim 0.1$, then $\sigma(\DE)$ and $\sigma(w)$
are dominated by the constant parameters of the mass function ($A_0$, $a_0$, and $b_0$). 
When the mass-function is held fixed, $\DE$ is most sensitive to the normalization (bias)
 of the mass--observable relation and its redshift evolution, whereas $w$ is more sensitive to the
redshift evolution of the variance.

We only consider individually self-calibrated cluster surveys. 
But as \cite{cun08,cun09} show, cross-calibration is a powerful tool to improve
knowledge of the mass--observable nuisance parameters and thereby tighten cosmological
constraints.  Strategies to optimize follow-up observations of cluster samples \citet{wu09} are relevant in this regard.  
The effects of using cross-calibrated cluster surveys would be to decrease the sensitivity 
to the mass--observable parameters, thereby increasing the relative sensitivity to the MF/B 
parameters.  Our results adopted a specific parameterization of the mass function based on the assumption of 
collisionless dynamics of dark matter, with no gas physics.  Baryonic effects can be large on the scale of 
clusters \citet{stanek09a}, and simulations exploring a broader range of baryon physics behavior are needed 
to ensure that the uncertainties in the mass function do not play a more significant role in limiting our knowledge of dark energy.

In conclusion, our results further illuminate challenges to precision cosmology with galaxy clusters.
Observers and simulators should focus on characterizing the form of the mass--observable relation, since 
having the correct parameterization is essential to avoid biases in derived cosmological constraints.
While we find that a mass function accuracy of about $10\%$ is a sub-dominant source of degradation of 
cosmological constraints, it still  contributes measurably, particularly for surveys where significant 
cross-calibration or targeted follow-up will be possible.

\acknowledgments

The authors would like to thank Roman Scoccimarro, Ravi Sheth, Jochen Weller and Andrew Zentner 
for useful discussions as well as Dragan Huterer, Eduardo Rozo, Risa Wechsler and Hao-Yi Wu for 
comments on the paper draft.
C.C. is supported by DOE OJI grant under contract DE-FG02-95ER40899.
A.E.E. acknowledges support from NSF AST-0708150.  


\bibliography{massfunction}

\begin{thebibliography}{63}
\expandafter\ifx\csname natexlab\endcsname\relax\def\natexlab#1{#1}\fi
\expandafter\ifx\csname bibnamefont\endcsname\relax
  \def\bibnamefont#1{#1}\fi
\expandafter\ifx\csname bibfnamefont\endcsname\relax
  \def\bibfnamefont#1{#1}\fi
\expandafter\ifx\csname citenamefont\endcsname\relax
  \def\citenamefont#1{#1}\fi
\expandafter\ifx\csname url\endcsname\relax
  \def\url#1{\texttt{#1}}\fi
\expandafter\ifx\csname urlprefix\endcsname\relax\def\urlprefix{URL }\fi
\providecommand{\bibinfo}[2]{#2}
\providecommand{\eprint}[2][]{\url{#2}}

\bibitem[{\citenamefont{{Cunha} et~al.}(2009)\citenamefont{{Cunha}, {Huterer},
  and {Frieman}}}]{cun09}
\bibinfo{author}{\bibfnamefont{C.}~\bibnamefont{{Cunha}}},
  \bibinfo{author}{\bibfnamefont{D.}~\bibnamefont{{Huterer}}},
  \bibnamefont{and} \bibinfo{author}{\bibfnamefont{J.~A.}
  \bibnamefont{{Frieman}}}, \bibinfo{journal}{ArXiv e-prints}
  (\bibinfo{year}{2009}), \eprint{0904.1589}.

\bibitem[{\citenamefont{{Sahl{\'e}n} et~al.}(2009)\citenamefont{{Sahl{\'e}n},
  {Viana}, {Liddle}, {Romer}, {Davidson}, {Hosmer}, {Lloyd-Davies}, {Sabirli},
  {Collins}, {Freeman} et~al.}}]{sah09}
\bibinfo{author}{\bibfnamefont{M.}~\bibnamefont{{Sahl{\'e}n}}},
  \bibinfo{author}{\bibfnamefont{P.~T.~P.} \bibnamefont{{Viana}}},
  \bibinfo{author}{\bibfnamefont{A.~R.} \bibnamefont{{Liddle}}},
  \bibinfo{author}{\bibfnamefont{A.~K.} \bibnamefont{{Romer}}},
  \bibinfo{author}{\bibfnamefont{M.}~\bibnamefont{{Davidson}}},
  \bibinfo{author}{\bibfnamefont{M.}~\bibnamefont{{Hosmer}}},
  \bibinfo{author}{\bibfnamefont{E.}~\bibnamefont{{Lloyd-Davies}}},
  \bibinfo{author}{\bibfnamefont{K.}~\bibnamefont{{Sabirli}}},
  \bibinfo{author}{\bibfnamefont{C.~A.} \bibnamefont{{Collins}}},
  \bibinfo{author}{\bibfnamefont{P.~E.} \bibnamefont{{Freeman}}},
  \bibnamefont{et~al.}, \bibinfo{journal}{\mnras}
  \textbf{\bibinfo{volume}{397}}, \bibinfo{pages}{577} (\bibinfo{year}{2009}),
  \eprint{0802.4462}.

\bibitem[{\citenamefont{{Voit}}(2005)}]{voi05}
\bibinfo{author}{\bibfnamefont{G.~M.} \bibnamefont{{Voit}}},
  \bibinfo{journal}{Reviews of Modern Physics} \textbf{\bibinfo{volume}{77}},
  \bibinfo{pages}{207} (\bibinfo{year}{2005}), \eprint{arXiv:astro-ph/0410173}.

\bibitem[{\citenamefont{{Battye} and {Weller}}(2003)}]{bat03}
\bibinfo{author}{\bibfnamefont{R.~A.} \bibnamefont{{Battye}}} \bibnamefont{and}
  \bibinfo{author}{\bibfnamefont{J.}~\bibnamefont{{Weller}}},
  \bibinfo{journal}{\prd} \textbf{\bibinfo{volume}{68}},
  \bibinfo{pages}{083506} (\bibinfo{year}{2003}),
  \eprint{arXiv:astro-ph/0305568}.

\bibitem[{\citenamefont{{Rosati} et~al.}(2002)\citenamefont{{Rosati},
  {Borgani}, and {Norman}}}]{ros02}
\bibinfo{author}{\bibfnamefont{P.}~\bibnamefont{{Rosati}}},
  \bibinfo{author}{\bibfnamefont{S.}~\bibnamefont{{Borgani}}},
  \bibnamefont{and} \bibinfo{author}{\bibfnamefont{C.}~\bibnamefont{{Norman}}},
  \bibinfo{journal}{\araa} \textbf{\bibinfo{volume}{40}}, \bibinfo{pages}{539}
  (\bibinfo{year}{2002}), \eprint{arXiv:astro-ph/0209035}.

\bibitem[{\citenamefont{{Haiman} et~al.}(2001)\citenamefont{{Haiman}, {Mohr},
  and {Holder}}}]{hai01}
\bibinfo{author}{\bibfnamefont{Z.}~\bibnamefont{{Haiman}}},
  \bibinfo{author}{\bibfnamefont{J.~J.} \bibnamefont{{Mohr}}},
  \bibnamefont{and} \bibinfo{author}{\bibfnamefont{G.~P.}
  \bibnamefont{{Holder}}}, \bibinfo{journal}{\apj}
  \textbf{\bibinfo{volume}{553}}, \bibinfo{pages}{545} (\bibinfo{year}{2001}),
  \eprint{arXiv:astro-ph/0002336}.

\bibitem[{\citenamefont{{Marian} and {Bernstein}}(2006)}]{mar06}
\bibinfo{author}{\bibfnamefont{L.}~\bibnamefont{{Marian}}} \bibnamefont{and}
  \bibinfo{author}{\bibfnamefont{G.~M.} \bibnamefont{{Bernstein}}},
  \bibinfo{journal}{\prd} \textbf{\bibinfo{volume}{73}},
  \bibinfo{pages}{123525} (\bibinfo{year}{2006}),
  \eprint{arXiv:astro-ph/0605746}.

\bibitem[{\citenamefont{{Mantz} et~al.}(2008)\citenamefont{{Mantz}, {Allen},
  {Ebeling}, and {Rapetti}}}]{man08}
\bibinfo{author}{\bibfnamefont{A.}~\bibnamefont{{Mantz}}},
  \bibinfo{author}{\bibfnamefont{S.~W.} \bibnamefont{{Allen}}},
  \bibinfo{author}{\bibfnamefont{H.}~\bibnamefont{{Ebeling}}},
  \bibnamefont{and}
  \bibinfo{author}{\bibfnamefont{D.}~\bibnamefont{{Rapetti}}},
  \bibinfo{journal}{\mnras} \textbf{\bibinfo{volume}{387}},
  \bibinfo{pages}{1179} (\bibinfo{year}{2008}), \eprint{0709.4294}.

\bibitem[{\citenamefont{{Vikhlinin} et~al.}(2009)\citenamefont{{Vikhlinin},
  {Kravtsov}, {Burenin}, {Ebeling}, {Forman}, {Hornstrup}, {Jones}, {Murray},
  {Nagai}, {Quintana} et~al.}}]{vik09}
\bibinfo{author}{\bibfnamefont{A.}~\bibnamefont{{Vikhlinin}}},
  \bibinfo{author}{\bibfnamefont{A.~V.} \bibnamefont{{Kravtsov}}},
  \bibinfo{author}{\bibfnamefont{R.~A.} \bibnamefont{{Burenin}}},
  \bibinfo{author}{\bibfnamefont{H.}~\bibnamefont{{Ebeling}}},
  \bibinfo{author}{\bibfnamefont{W.~R.} \bibnamefont{{Forman}}},
  \bibinfo{author}{\bibfnamefont{A.}~\bibnamefont{{Hornstrup}}},
  \bibinfo{author}{\bibfnamefont{C.}~\bibnamefont{{Jones}}},
  \bibinfo{author}{\bibfnamefont{S.~S.} \bibnamefont{{Murray}}},
  \bibinfo{author}{\bibfnamefont{D.}~\bibnamefont{{Nagai}}},
  \bibinfo{author}{\bibfnamefont{H.}~\bibnamefont{{Quintana}}},
  \bibnamefont{et~al.}, \bibinfo{journal}{\apj} \textbf{\bibinfo{volume}{692}},
  \bibinfo{pages}{1060} (\bibinfo{year}{2009}), \eprint{0812.2720}.

\bibitem[{\citenamefont{{Henry} et~al.}(2009)\citenamefont{{Henry}, {Evrard},
  {Hoekstra}, {Babul}, and {Mahdavi}}}]{hen09}
\bibinfo{author}{\bibfnamefont{J.~P.} \bibnamefont{{Henry}}},
  \bibinfo{author}{\bibfnamefont{A.~E.} \bibnamefont{{Evrard}}},
  \bibinfo{author}{\bibfnamefont{H.}~\bibnamefont{{Hoekstra}}},
  \bibinfo{author}{\bibfnamefont{A.}~\bibnamefont{{Babul}}}, \bibnamefont{and}
  \bibinfo{author}{\bibfnamefont{A.}~\bibnamefont{{Mahdavi}}},
  \bibinfo{journal}{\apj} \textbf{\bibinfo{volume}{691}}, \bibinfo{pages}{1307}
  (\bibinfo{year}{2009}), \eprint{0809.3832}.

\bibitem[{\citenamefont{{Gladders} et~al.}(2007)\citenamefont{{Gladders},
  {Yee}, {Majumdar}, {Barrientos}, {Hoekstra}, {Hall}, and {Infante}}}]{gla07}
\bibinfo{author}{\bibfnamefont{M.~D.} \bibnamefont{{Gladders}}},
  \bibinfo{author}{\bibfnamefont{H.~K.~C.} \bibnamefont{{Yee}}},
  \bibinfo{author}{\bibfnamefont{S.}~\bibnamefont{{Majumdar}}},
  \bibinfo{author}{\bibfnamefont{L.~F.} \bibnamefont{{Barrientos}}},
  \bibinfo{author}{\bibfnamefont{H.}~\bibnamefont{{Hoekstra}}},
  \bibinfo{author}{\bibfnamefont{P.~B.} \bibnamefont{{Hall}}},
  \bibnamefont{and}
  \bibinfo{author}{\bibfnamefont{L.}~\bibnamefont{{Infante}}},
  \bibinfo{journal}{\apj} \textbf{\bibinfo{volume}{655}}, \bibinfo{pages}{128}
  (\bibinfo{year}{2007}), \eprint{arXiv:astro-ph/0603588}.

\bibitem[{\citenamefont{{Rozo} et~al.}(2009)\citenamefont{{Rozo}, {Wechsler},
  {Rykoff}, {Annis}, {Becker}, {Evrard}, {Frieman}, {Hansen}, {Hao}, {Johnston}
  et~al.}}]{roz09}
\bibinfo{author}{\bibfnamefont{E.}~\bibnamefont{{Rozo}}},
  \bibinfo{author}{\bibfnamefont{R.~H.} \bibnamefont{{Wechsler}}},
  \bibinfo{author}{\bibfnamefont{E.~S.} \bibnamefont{{Rykoff}}},
  \bibinfo{author}{\bibfnamefont{J.~T.} \bibnamefont{{Annis}}},
  \bibinfo{author}{\bibfnamefont{M.~R.} \bibnamefont{{Becker}}},
  \bibinfo{author}{\bibfnamefont{A.~E.} \bibnamefont{{Evrard}}},
  \bibinfo{author}{\bibfnamefont{J.~A.} \bibnamefont{{Frieman}}},
  \bibinfo{author}{\bibfnamefont{S.~M.} \bibnamefont{{Hansen}}},
  \bibinfo{author}{\bibfnamefont{J.}~\bibnamefont{{Hao}}},
  \bibinfo{author}{\bibfnamefont{D.~E.} \bibnamefont{{Johnston}}},
  \bibnamefont{et~al.}, \bibinfo{journal}{ArXiv e-prints}
  (\bibinfo{year}{2009}), \eprint{0902.3702}.

\bibitem[{\citenamefont{{Levine} et~al.}(2002)\citenamefont{{Levine}, {Schulz},
  and {White}}}]{lev02}
\bibinfo{author}{\bibfnamefont{E.~S.} \bibnamefont{{Levine}}},
  \bibinfo{author}{\bibfnamefont{A.~E.} \bibnamefont{{Schulz}}},
  \bibnamefont{and} \bibinfo{author}{\bibfnamefont{M.}~\bibnamefont{{White}}},
  \bibinfo{journal}{\apj} \textbf{\bibinfo{volume}{577}}, \bibinfo{pages}{569}
  (\bibinfo{year}{2002}), \eprint{arXiv:astro-ph/0204273}.

\bibitem[{\citenamefont{{Majumdar} and {Mohr}}(2003)}]{maj03}
\bibinfo{author}{\bibfnamefont{S.}~\bibnamefont{{Majumdar}}} \bibnamefont{and}
  \bibinfo{author}{\bibfnamefont{J.~J.} \bibnamefont{{Mohr}}},
  \bibinfo{journal}{\apj} \textbf{\bibinfo{volume}{585}}, \bibinfo{pages}{603}
  (\bibinfo{year}{2003}), \eprint{arXiv:astro-ph/0208002}.

\bibitem[{\citenamefont{{Majumdar} and {Mohr}}(2004)}]{maj04}
\bibinfo{author}{\bibfnamefont{S.}~\bibnamefont{{Majumdar}}} \bibnamefont{and}
  \bibinfo{author}{\bibfnamefont{J.~J.} \bibnamefont{{Mohr}}},
  \bibinfo{journal}{\apj} \textbf{\bibinfo{volume}{613}}, \bibinfo{pages}{41}
  (\bibinfo{year}{2004}), \eprint{arXiv:astro-ph/0305341}.

\bibitem[{\citenamefont{{Lima} and {Hu}}(2004)}]{lim04}
\bibinfo{author}{\bibfnamefont{M.}~\bibnamefont{{Lima}}} \bibnamefont{and}
  \bibinfo{author}{\bibfnamefont{W.}~\bibnamefont{{Hu}}},
  \bibinfo{journal}{\prd} \textbf{\bibinfo{volume}{70}},
  \bibinfo{pages}{043504} (\bibinfo{year}{2004}),
  \eprint{arXiv:astro-ph/0401559}.

\bibitem[{\citenamefont{{Lima} and {Hu}}(2005)}]{lim05}
\bibinfo{author}{\bibfnamefont{M.}~\bibnamefont{{Lima}}} \bibnamefont{and}
  \bibinfo{author}{\bibfnamefont{W.}~\bibnamefont{{Hu}}},
  \bibinfo{journal}{\prd} \textbf{\bibinfo{volume}{72}},
  \bibinfo{pages}{043006} (\bibinfo{year}{2005}),
  \eprint{arXiv:astro-ph/0503363}.

\bibitem[{\citenamefont{{Lima} and {Hu}}(2007)}]{lim07}
\bibinfo{author}{\bibfnamefont{M.}~\bibnamefont{{Lima}}} \bibnamefont{and}
  \bibinfo{author}{\bibfnamefont{W.}~\bibnamefont{{Hu}}},
  \bibinfo{journal}{\prd} \textbf{\bibinfo{volume}{76}},
  \bibinfo{pages}{123013} (\bibinfo{year}{2007}), \eprint{arXiv:0709.2871}.

\bibitem[{\citenamefont{Cunha}(2009)}]{cun08}
\bibinfo{author}{\bibfnamefont{C.~E.} \bibnamefont{Cunha}},
  \bibinfo{journal}{\prd} \textbf{\bibinfo{volume}{79}},
  \bibinfo{pages}{063009} (\bibinfo{year}{2009}).

\bibitem[{\citenamefont{Shaw et~al.}(2009)\citenamefont{Shaw, Holder, and
  Dudley}}]{shaw09}
\bibinfo{author}{\bibfnamefont{L.}~\bibnamefont{Shaw}},
  \bibinfo{author}{\bibfnamefont{G.}~\bibnamefont{Holder}}, \bibnamefont{and}
  \bibinfo{author}{\bibfnamefont{J.}~\bibnamefont{Dudley}},
  \bibinfo{journal}{in preparation}  (\bibinfo{year}{2009}).

\bibitem[{\citenamefont{{Arnaud} et~al.}(2005)\citenamefont{{Arnaud},
  {Pointecouteau}, and {Pratt}}}]{arnaud05}
\bibinfo{author}{\bibfnamefont{M.}~\bibnamefont{{Arnaud}}},
  \bibinfo{author}{\bibfnamefont{E.}~\bibnamefont{{Pointecouteau}}},
  \bibnamefont{and} \bibinfo{author}{\bibfnamefont{G.~W.}
  \bibnamefont{{Pratt}}}, \bibinfo{journal}{\aap}
  \textbf{\bibinfo{volume}{441}}, \bibinfo{pages}{893} (\bibinfo{year}{2005}),
  \eprint{arXiv:astro-ph/0502210}.

\bibitem[{\citenamefont{{Maughan} et~al.}(2006)\citenamefont{{Maughan},
  {Jones}, {Ebeling}, and {Scharf}}}]{maughan06}
\bibinfo{author}{\bibfnamefont{B.~J.} \bibnamefont{{Maughan}}},
  \bibinfo{author}{\bibfnamefont{L.~R.} \bibnamefont{{Jones}}},
  \bibinfo{author}{\bibfnamefont{H.}~\bibnamefont{{Ebeling}}},
  \bibnamefont{and} \bibinfo{author}{\bibfnamefont{C.}~\bibnamefont{{Scharf}}},
  \bibinfo{journal}{\mnras} \textbf{\bibinfo{volume}{365}},
  \bibinfo{pages}{509} (\bibinfo{year}{2006}), \eprint{arXiv:astro-ph/0503455}.

\bibitem[{\citenamefont{{Vikhlinin} et~al.}(2006)\citenamefont{{Vikhlinin},
  {Kravtsov}, {Forman}, {Jones}, {Markevitch}, {Murray}, and {Van
  Speybroeck}}}]{vikhlinin06}
\bibinfo{author}{\bibfnamefont{A.}~\bibnamefont{{Vikhlinin}}},
  \bibinfo{author}{\bibfnamefont{A.}~\bibnamefont{{Kravtsov}}},
  \bibinfo{author}{\bibfnamefont{W.}~\bibnamefont{{Forman}}},
  \bibinfo{author}{\bibfnamefont{C.}~\bibnamefont{{Jones}}},
  \bibinfo{author}{\bibfnamefont{M.}~\bibnamefont{{Markevitch}}},
  \bibinfo{author}{\bibfnamefont{S.~S.} \bibnamefont{{Murray}}},
  \bibnamefont{and} \bibinfo{author}{\bibfnamefont{L.}~\bibnamefont{{Van
  Speybroeck}}}, \bibinfo{journal}{\apj} \textbf{\bibinfo{volume}{640}},
  \bibinfo{pages}{691} (\bibinfo{year}{2006}), \eprint{arXiv:astro-ph/0507092}.

\bibitem[{\citenamefont{{Morandi} et~al.}(2007)\citenamefont{{Morandi},
  {Ettori}, and {Moscardini}}}]{morandi07}
\bibinfo{author}{\bibfnamefont{A.}~\bibnamefont{{Morandi}}},
  \bibinfo{author}{\bibfnamefont{S.}~\bibnamefont{{Ettori}}}, \bibnamefont{and}
  \bibinfo{author}{\bibfnamefont{L.}~\bibnamefont{{Moscardini}}},
  \bibinfo{journal}{\mnras} \textbf{\bibinfo{volume}{379}},
  \bibinfo{pages}{518} (\bibinfo{year}{2007}), \eprint{0704.2678}.

\bibitem[{\citenamefont{{Bonamente} et~al.}(2008)\citenamefont{{Bonamente},
  {Joy}, {LaRoque}, {Carlstrom}, {Nagai}, and {Marrone}}}]{bonamente08}
\bibinfo{author}{\bibfnamefont{M.}~\bibnamefont{{Bonamente}}},
  \bibinfo{author}{\bibfnamefont{M.}~\bibnamefont{{Joy}}},
  \bibinfo{author}{\bibfnamefont{S.~J.} \bibnamefont{{LaRoque}}},
  \bibinfo{author}{\bibfnamefont{J.~E.} \bibnamefont{{Carlstrom}}},
  \bibinfo{author}{\bibfnamefont{D.}~\bibnamefont{{Nagai}}}, \bibnamefont{and}
  \bibinfo{author}{\bibfnamefont{D.~P.} \bibnamefont{{Marrone}}},
  \bibinfo{journal}{\apj} \textbf{\bibinfo{volume}{675}}, \bibinfo{pages}{106}
  (\bibinfo{year}{2008}), \eprint{0708.0815}.

\bibitem[{\citenamefont{{Zhang} et~al.}(2008)\citenamefont{{Zhang},
  {Finoguenov}, {B{\"o}hringer}, {Kneib}, {Smith}, {Kneissl}, {Okabe}, and
  {Dahle}}}]{zhang08}
\bibinfo{author}{\bibfnamefont{Y.-Y.} \bibnamefont{{Zhang}}},
  \bibinfo{author}{\bibfnamefont{A.}~\bibnamefont{{Finoguenov}}},
  \bibinfo{author}{\bibfnamefont{H.}~\bibnamefont{{B{\"o}hringer}}},
  \bibinfo{author}{\bibfnamefont{J.-P.} \bibnamefont{{Kneib}}},
  \bibinfo{author}{\bibfnamefont{G.~P.} \bibnamefont{{Smith}}},
  \bibinfo{author}{\bibfnamefont{R.}~\bibnamefont{{Kneissl}}},
  \bibinfo{author}{\bibfnamefont{N.}~\bibnamefont{{Okabe}}}, \bibnamefont{and}
  \bibinfo{author}{\bibfnamefont{H.}~\bibnamefont{{Dahle}}},
  \bibinfo{journal}{\aap} \textbf{\bibinfo{volume}{482}}, \bibinfo{pages}{451}
  (\bibinfo{year}{2008}), \eprint{0802.0770}.

\bibitem[{\citenamefont{{Pratt} et~al.}(2009)\citenamefont{{Pratt}, {Croston},
  {Arnaud}, and {B{\"o}hringer}}}]{pratt09}
\bibinfo{author}{\bibfnamefont{G.~W.} \bibnamefont{{Pratt}}},
  \bibinfo{author}{\bibfnamefont{J.~H.} \bibnamefont{{Croston}}},
  \bibinfo{author}{\bibfnamefont{M.}~\bibnamefont{{Arnaud}}}, \bibnamefont{and}
  \bibinfo{author}{\bibfnamefont{H.}~\bibnamefont{{B{\"o}hringer}}},
  \bibinfo{journal}{\aap} \textbf{\bibinfo{volume}{498}}, \bibinfo{pages}{361}
  (\bibinfo{year}{2009}), \eprint{0809.3784}.

\bibitem[{\citenamefont{{Bialek} et~al.}(2001)\citenamefont{{Bialek}, {Evrard},
  and {Mohr}}}]{bialek01}
\bibinfo{author}{\bibfnamefont{J.~J.} \bibnamefont{{Bialek}}},
  \bibinfo{author}{\bibfnamefont{A.~E.} \bibnamefont{{Evrard}}},
  \bibnamefont{and} \bibinfo{author}{\bibfnamefont{J.~J.}
  \bibnamefont{{Mohr}}}, \bibinfo{journal}{\apj}
  \textbf{\bibinfo{volume}{555}}, \bibinfo{pages}{597} (\bibinfo{year}{2001}),
  \eprint{arXiv:astro-ph/0010584}.

\bibitem[{\citenamefont{{Borgani} et~al.}(2004)\citenamefont{{Borgani},
  {Murante}, {Springel}, {Diaferio}, {Dolag}, {Moscardini}, {Tormen},
  {Tornatore}, and {Tozzi}}}]{borgani04}
\bibinfo{author}{\bibfnamefont{S.}~\bibnamefont{{Borgani}}},
  \bibinfo{author}{\bibfnamefont{G.}~\bibnamefont{{Murante}}},
  \bibinfo{author}{\bibfnamefont{V.}~\bibnamefont{{Springel}}},
  \bibinfo{author}{\bibfnamefont{A.}~\bibnamefont{{Diaferio}}},
  \bibinfo{author}{\bibfnamefont{K.}~\bibnamefont{{Dolag}}},
  \bibinfo{author}{\bibfnamefont{L.}~\bibnamefont{{Moscardini}}},
  \bibinfo{author}{\bibfnamefont{G.}~\bibnamefont{{Tormen}}},
  \bibinfo{author}{\bibfnamefont{L.}~\bibnamefont{{Tornatore}}},
  \bibnamefont{and} \bibinfo{author}{\bibfnamefont{P.}~\bibnamefont{{Tozzi}}},
  \bibinfo{journal}{\mnras} \textbf{\bibinfo{volume}{348}},
  \bibinfo{pages}{1078} (\bibinfo{year}{2004}),
  \eprint{arXiv:astro-ph/0310794}.

\bibitem[{\citenamefont{{da Silva} et~al.}(2004)\citenamefont{{da Silva},
  {Kay}, {Liddle}, and {Thomas}}}]{dasilva04}
\bibinfo{author}{\bibfnamefont{A.~C.} \bibnamefont{{da Silva}}},
  \bibinfo{author}{\bibfnamefont{S.~T.} \bibnamefont{{Kay}}},
  \bibinfo{author}{\bibfnamefont{A.~R.} \bibnamefont{{Liddle}}},
  \bibnamefont{and} \bibinfo{author}{\bibfnamefont{P.~A.}
  \bibnamefont{{Thomas}}}, \bibinfo{journal}{\mnras}
  \textbf{\bibinfo{volume}{348}}, \bibinfo{pages}{1401} (\bibinfo{year}{2004}),
  \eprint{arXiv:astro-ph/0308074}.

\bibitem[{\citenamefont{{Kravtsov} et~al.}(2006)\citenamefont{{Kravtsov},
  {Vikhlinin}, and {Nagai}}}]{kravtsov06}
\bibinfo{author}{\bibfnamefont{A.~V.} \bibnamefont{{Kravtsov}}},
  \bibinfo{author}{\bibfnamefont{A.}~\bibnamefont{{Vikhlinin}}},
  \bibnamefont{and} \bibinfo{author}{\bibfnamefont{D.}~\bibnamefont{{Nagai}}},
  \bibinfo{journal}{\apj} \textbf{\bibinfo{volume}{650}}, \bibinfo{pages}{128}
  (\bibinfo{year}{2006}), \eprint{arXiv:astro-ph/0603205}.

\bibitem[{\citenamefont{{Ascasibar} et~al.}(2006)\citenamefont{{Ascasibar},
  {Sevilla}, {Yepes}, {M{\"u}ller}, and {Gottl{\"o}ber}}}]{ascasibar06}
\bibinfo{author}{\bibfnamefont{Y.}~\bibnamefont{{Ascasibar}}},
  \bibinfo{author}{\bibfnamefont{R.}~\bibnamefont{{Sevilla}}},
  \bibinfo{author}{\bibfnamefont{G.}~\bibnamefont{{Yepes}}},
  \bibinfo{author}{\bibfnamefont{V.}~\bibnamefont{{M{\"u}ller}}},
  \bibnamefont{and}
  \bibinfo{author}{\bibfnamefont{S.}~\bibnamefont{{Gottl{\"o}ber}}},
  \bibinfo{journal}{\mnras} \textbf{\bibinfo{volume}{371}},
  \bibinfo{pages}{193} (\bibinfo{year}{2006}), \eprint{arXiv:astro-ph/0605720}.

\bibitem[{\citenamefont{{Muanwong} et~al.}(2006)\citenamefont{{Muanwong},
  {Kay}, and {Thomas}}}]{muanwong06}
\bibinfo{author}{\bibfnamefont{O.}~\bibnamefont{{Muanwong}}},
  \bibinfo{author}{\bibfnamefont{S.~T.} \bibnamefont{{Kay}}}, \bibnamefont{and}
  \bibinfo{author}{\bibfnamefont{P.~A.} \bibnamefont{{Thomas}}},
  \bibinfo{journal}{\apj} \textbf{\bibinfo{volume}{649}}, \bibinfo{pages}{640}
  (\bibinfo{year}{2006}), \eprint{arXiv:astro-ph/0509803}.

\bibitem[{\citenamefont{{Puchwein} et~al.}(2008)\citenamefont{{Puchwein},
  {Sijacki}, and {Springel}}}]{puchwein08}
\bibinfo{author}{\bibfnamefont{E.}~\bibnamefont{{Puchwein}}},
  \bibinfo{author}{\bibfnamefont{D.}~\bibnamefont{{Sijacki}}},
  \bibnamefont{and}
  \bibinfo{author}{\bibfnamefont{V.}~\bibnamefont{{Springel}}},
  \bibinfo{journal}{\apjl} \textbf{\bibinfo{volume}{687}}, \bibinfo{pages}{L53}
  (\bibinfo{year}{2008}), \eprint{0808.0494}.

\bibitem[{\citenamefont{{Aghanim} et~al.}(2009)\citenamefont{{Aghanim}, {da
  Silva}, and {Nunes}}}]{aghanim09}
\bibinfo{author}{\bibfnamefont{N.}~\bibnamefont{{Aghanim}}},
  \bibinfo{author}{\bibfnamefont{A.~C.} \bibnamefont{{da Silva}}},
  \bibnamefont{and} \bibinfo{author}{\bibfnamefont{N.~J.}
  \bibnamefont{{Nunes}}}, \bibinfo{journal}{\aap}
  \textbf{\bibinfo{volume}{496}}, \bibinfo{pages}{637} (\bibinfo{year}{2009}),
  \eprint{0808.0385}.

\bibitem[{\citenamefont{Stanek et~al.}(2009)\citenamefont{Stanek, Rasia,
  Evrard, Pearce, and Gazzola}}]{stanek09b}
\bibinfo{author}{\bibfnamefont{R.}~\bibnamefont{Stanek}},
  \bibinfo{author}{\bibfnamefont{E.}~\bibnamefont{Rasia}},
  \bibinfo{author}{\bibfnamefont{A.}~\bibnamefont{Evrard}},
  \bibinfo{author}{\bibfnamefont{F.}~\bibnamefont{Pearce}}, \bibnamefont{and}
  \bibinfo{author}{\bibfnamefont{L.}~\bibnamefont{Gazzola}},
  \bibinfo{journal}{in preparation}  (\bibinfo{year}{2009}).

\bibitem[{\citenamefont{{Press} and {Schechter}}(1974)}]{pre74}
\bibinfo{author}{\bibfnamefont{W.~H.} \bibnamefont{{Press}}} \bibnamefont{and}
  \bibinfo{author}{\bibfnamefont{P.}~\bibnamefont{{Schechter}}},
  \bibinfo{journal}{\apj} \textbf{\bibinfo{volume}{187}}, \bibinfo{pages}{425}
  (\bibinfo{year}{1974}).

\bibitem[{\citenamefont{{Sheth} and {Tormen}}(1999)}]{she99}
\bibinfo{author}{\bibfnamefont{R.~K.} \bibnamefont{{Sheth}}} \bibnamefont{and}
  \bibinfo{author}{\bibfnamefont{G.}~\bibnamefont{{Tormen}}},
  \bibinfo{journal}{\mnras} \textbf{\bibinfo{volume}{308}},
  \bibinfo{pages}{119} (\bibinfo{year}{1999}), \eprint{arXiv:astro-ph/9901122}.

\bibitem[{\citenamefont{{Jenkins} et~al.}(2001)\citenamefont{{Jenkins},
  {Frenk}, {White}, {Colberg}, {Cole}, {Evrard}, {Couchman}, and
  {Yoshida}}}]{jen01}
\bibinfo{author}{\bibfnamefont{A.}~\bibnamefont{{Jenkins}}},
  \bibinfo{author}{\bibfnamefont{C.~S.} \bibnamefont{{Frenk}}},
  \bibinfo{author}{\bibfnamefont{S.~D.~M.} \bibnamefont{{White}}},
  \bibinfo{author}{\bibfnamefont{J.~M.} \bibnamefont{{Colberg}}},
  \bibinfo{author}{\bibfnamefont{S.}~\bibnamefont{{Cole}}},
  \bibinfo{author}{\bibfnamefont{A.~E.} \bibnamefont{{Evrard}}},
  \bibinfo{author}{\bibfnamefont{H.~M.~P.} \bibnamefont{{Couchman}}},
  \bibnamefont{and}
  \bibinfo{author}{\bibfnamefont{N.}~\bibnamefont{{Yoshida}}},
  \bibinfo{journal}{\mnras} \textbf{\bibinfo{volume}{321}},
  \bibinfo{pages}{372} (\bibinfo{year}{2001}), \eprint{arXiv:astro-ph/0005260}.

\bibitem[{\citenamefont{{Evrard} et~al.}(2002)\citenamefont{{Evrard},
  {MacFarland}, {Couchman}, {Colberg}, {Yoshida}, {White}, {Jenkins}, {Frenk},
  {Pearce}, {Peacock} et~al.}}]{evr02}
\bibinfo{author}{\bibfnamefont{A.~E.} \bibnamefont{{Evrard}}},
  \bibinfo{author}{\bibfnamefont{T.~J.} \bibnamefont{{MacFarland}}},
  \bibinfo{author}{\bibfnamefont{H.~M.~P.} \bibnamefont{{Couchman}}},
  \bibinfo{author}{\bibfnamefont{J.~M.} \bibnamefont{{Colberg}}},
  \bibinfo{author}{\bibfnamefont{N.}~\bibnamefont{{Yoshida}}},
  \bibinfo{author}{\bibfnamefont{S.~D.~M.} \bibnamefont{{White}}},
  \bibinfo{author}{\bibfnamefont{A.}~\bibnamefont{{Jenkins}}},
  \bibinfo{author}{\bibfnamefont{C.~S.} \bibnamefont{{Frenk}}},
  \bibinfo{author}{\bibfnamefont{F.~R.} \bibnamefont{{Pearce}}},
  \bibinfo{author}{\bibfnamefont{J.~A.} \bibnamefont{{Peacock}}},
  \bibnamefont{et~al.}, \bibinfo{journal}{\apj} \textbf{\bibinfo{volume}{573}},
  \bibinfo{pages}{7} (\bibinfo{year}{2002}), \eprint{arXiv:astro-ph/0110246}.

\bibitem[{\citenamefont{{Warren} et~al.}(2006)\citenamefont{{Warren},
  {Abazajian}, {Holz}, and {Teodoro}}}]{war06}
\bibinfo{author}{\bibfnamefont{M.~S.} \bibnamefont{{Warren}}},
  \bibinfo{author}{\bibfnamefont{K.}~\bibnamefont{{Abazajian}}},
  \bibinfo{author}{\bibfnamefont{D.~E.} \bibnamefont{{Holz}}},
  \bibnamefont{and}
  \bibinfo{author}{\bibfnamefont{L.}~\bibnamefont{{Teodoro}}},
  \bibinfo{journal}{\apj} \textbf{\bibinfo{volume}{646}}, \bibinfo{pages}{881}
  (\bibinfo{year}{2006}), \eprint{arXiv:astro-ph/0506395}.

\bibitem[{\citenamefont{{Tinker} et~al.}(2008)\citenamefont{{Tinker},
  {Kravtsov}, {Klypin}, {Abazajian}, {Warren}, {Yepes}, {Gottlober}, and
  {Holz}}}]{tin08}
\bibinfo{author}{\bibfnamefont{J.~L.} \bibnamefont{{Tinker}}},
  \bibinfo{author}{\bibfnamefont{A.~V.} \bibnamefont{{Kravtsov}}},
  \bibinfo{author}{\bibfnamefont{A.}~\bibnamefont{{Klypin}}},
  \bibinfo{author}{\bibfnamefont{K.}~\bibnamefont{{Abazajian}}},
  \bibinfo{author}{\bibfnamefont{M.~S.} \bibnamefont{{Warren}}},
  \bibinfo{author}{\bibfnamefont{G.}~\bibnamefont{{Yepes}}},
  \bibinfo{author}{\bibfnamefont{S.}~\bibnamefont{{Gottlober}}},
  \bibnamefont{and} \bibinfo{author}{\bibfnamefont{D.~E.}
  \bibnamefont{{Holz}}}, \bibinfo{journal}{ArXiv e-prints}
  \textbf{\bibinfo{volume}{803}} (\bibinfo{year}{2008}), \eprint{0803.2706}.

\bibitem[{\citenamefont{{Crocce} et~al.}(2009)\citenamefont{{Crocce},
  {Fosalba}, {Castander}, and {Gaztanaga}}}]{cro09}
\bibinfo{author}{\bibfnamefont{M.}~\bibnamefont{{Crocce}}},
  \bibinfo{author}{\bibfnamefont{P.}~\bibnamefont{{Fosalba}}},
  \bibinfo{author}{\bibfnamefont{F.~J.} \bibnamefont{{Castander}}},
  \bibnamefont{and}
  \bibinfo{author}{\bibfnamefont{E.}~\bibnamefont{{Gaztanaga}}},
  \bibinfo{journal}{ArXiv e-prints}  (\bibinfo{year}{2009}),
  \eprint{0907.0019}.

\bibitem[{\citenamefont{{Cole} and {Lacey}}(1996)}]{laceyCole96}
\bibinfo{author}{\bibfnamefont{S.}~\bibnamefont{{Cole}}} \bibnamefont{and}
  \bibinfo{author}{\bibfnamefont{C.}~\bibnamefont{{Lacey}}},
  \bibinfo{journal}{\mnras} \textbf{\bibinfo{volume}{281}},
  \bibinfo{pages}{716} (\bibinfo{year}{1996}), \eprint{arXiv:astro-ph/9510147}.

\bibitem[{\citenamefont{{White}}(2002)}]{white02}
\bibinfo{author}{\bibfnamefont{M.}~\bibnamefont{{White}}},
  \bibinfo{journal}{\apjs} \textbf{\bibinfo{volume}{143}}, \bibinfo{pages}{241}
  (\bibinfo{year}{2002}), \eprint{astro-ph/0207185}.

\bibitem[{\citenamefont{{Luki{\'c}} et~al.}(2009)\citenamefont{{Luki{\'c}},
  {Reed}, {Habib}, and {Heitmann}}}]{lukic09}
\bibinfo{author}{\bibfnamefont{Z.}~\bibnamefont{{Luki{\'c}}}},
  \bibinfo{author}{\bibfnamefont{D.}~\bibnamefont{{Reed}}},
  \bibinfo{author}{\bibfnamefont{S.}~\bibnamefont{{Habib}}}, \bibnamefont{and}
  \bibinfo{author}{\bibfnamefont{K.}~\bibnamefont{{Heitmann}}},
  \bibinfo{journal}{\apj} \textbf{\bibinfo{volume}{692}}, \bibinfo{pages}{217}
  (\bibinfo{year}{2009}), \eprint{0803.3624}.

\bibitem[{\citenamefont{{Stanek} et~al.}(2009)\citenamefont{{Stanek}, {Rudd},
  and {Evrard}}}]{stanek09a}
\bibinfo{author}{\bibfnamefont{R.}~\bibnamefont{{Stanek}}},
  \bibinfo{author}{\bibfnamefont{D.}~\bibnamefont{{Rudd}}}, \bibnamefont{and}
  \bibinfo{author}{\bibfnamefont{A.~E.} \bibnamefont{{Evrard}}},
  \bibinfo{journal}{\mnras} \textbf{\bibinfo{volume}{394}},
  \bibinfo{pages}{L11} (\bibinfo{year}{2009}), \eprint{0809.2805}.

\bibitem[{\citenamefont{{Lo Verde} et~al.}(2008)\citenamefont{{Lo Verde},
  {Miller}, {Shandera}, and {Verde}}}]{lov08}
\bibinfo{author}{\bibfnamefont{M.}~\bibnamefont{{Lo Verde}}},
  \bibinfo{author}{\bibfnamefont{A.}~\bibnamefont{{Miller}}},
  \bibinfo{author}{\bibfnamefont{S.}~\bibnamefont{{Shandera}}},
  \bibnamefont{and} \bibinfo{author}{\bibfnamefont{L.}~\bibnamefont{{Verde}}},
  \bibinfo{journal}{Journal of Cosmology and Astro-Particle Physics}
  \textbf{\bibinfo{volume}{4}}, \bibinfo{pages}{14} (\bibinfo{year}{2008}),
  \eprint{0711.4126}.

\bibitem[{\citenamefont{{Dalal} et~al.}(2008)\citenamefont{{Dalal}, {Dor{\'e}},
  {Huterer}, and {Shirokov}}}]{dal08}
\bibinfo{author}{\bibfnamefont{N.}~\bibnamefont{{Dalal}}},
  \bibinfo{author}{\bibfnamefont{O.}~\bibnamefont{{Dor{\'e}}}},
  \bibinfo{author}{\bibfnamefont{D.}~\bibnamefont{{Huterer}}},
  \bibnamefont{and}
  \bibinfo{author}{\bibfnamefont{A.}~\bibnamefont{{Shirokov}}},
  \bibinfo{journal}{\prd} \textbf{\bibinfo{volume}{77}},
  \bibinfo{pages}{123514} (\bibinfo{year}{2008}), \eprint{0710.4560}.

\bibitem[{\citenamefont{{Grossi} et~al.}(2009)\citenamefont{{Grossi}, {Verde},
  {Carbone}, {Dolag}, {Branchini}, {Iannuzzi}, {Matarrese}, and
  {Moscardini}}}]{gro09}
\bibinfo{author}{\bibfnamefont{M.}~\bibnamefont{{Grossi}}},
  \bibinfo{author}{\bibfnamefont{L.}~\bibnamefont{{Verde}}},
  \bibinfo{author}{\bibfnamefont{C.}~\bibnamefont{{Carbone}}},
  \bibinfo{author}{\bibfnamefont{K.}~\bibnamefont{{Dolag}}},
  \bibinfo{author}{\bibfnamefont{E.}~\bibnamefont{{Branchini}}},
  \bibinfo{author}{\bibfnamefont{F.}~\bibnamefont{{Iannuzzi}}},
  \bibinfo{author}{\bibfnamefont{S.}~\bibnamefont{{Matarrese}}},
  \bibnamefont{and}
  \bibinfo{author}{\bibfnamefont{L.}~\bibnamefont{{Moscardini}}},
  \bibinfo{journal}{ArXiv e-prints}  (\bibinfo{year}{2009}),
  \eprint{0902.2013}.

\bibitem[{\citenamefont{Wu et~al.}(2009)\citenamefont{Wu, Zentner, and
  Wechsler}}]{wu09b}
\bibinfo{author}{\bibfnamefont{H.-Y.} \bibnamefont{Wu}},
  \bibinfo{author}{\bibfnamefont{A.~R.} \bibnamefont{Zentner}},
  \bibnamefont{and} \bibinfo{author}{\bibfnamefont{R.~H.}
  \bibnamefont{Wechsler}}, \emph{\bibinfo{title}{The impact of theoretical
  uncertainties in the halo mass function and halo bias on precision
  cosmology}} (\bibinfo{year}{2009}),
  \urlprefix\url{http://www.citebase.org/abstract?id=oai:arXiv.org:0910.3668}.

\bibitem[{\citenamefont{{Hu} and {Kravtsov}}(2003)}]{hu03}
\bibinfo{author}{\bibfnamefont{W.}~\bibnamefont{{Hu}}} \bibnamefont{and}
  \bibinfo{author}{\bibfnamefont{A.~V.} \bibnamefont{{Kravtsov}}},
  \bibinfo{journal}{\apj} \textbf{\bibinfo{volume}{584}}, \bibinfo{pages}{702}
  (\bibinfo{year}{2003}), \eprint{arXiv:astro-ph/0203169}.

\bibitem[{\citenamefont{{Hu} and {Cohn}}(2006)}]{hu06}
\bibinfo{author}{\bibfnamefont{W.}~\bibnamefont{{Hu}}} \bibnamefont{and}
  \bibinfo{author}{\bibfnamefont{J.~D.} \bibnamefont{{Cohn}}},
  \bibinfo{journal}{\prd} \textbf{\bibinfo{volume}{73}},
  \bibinfo{pages}{067301} (\bibinfo{year}{2006}),
  \eprint{arXiv:astro-ph/0602147}.

\bibitem[{\citenamefont{{Holder} et~al.}(2001)\citenamefont{{Holder}, {Haiman},
  and {Mohr}}}]{hol01}
\bibinfo{author}{\bibfnamefont{G.}~\bibnamefont{{Holder}}},
  \bibinfo{author}{\bibfnamefont{Z.}~\bibnamefont{{Haiman}}}, \bibnamefont{and}
  \bibinfo{author}{\bibfnamefont{J.~J.} \bibnamefont{{Mohr}}},
  \bibinfo{journal}{\apjl} \textbf{\bibinfo{volume}{560}},
  \bibinfo{pages}{L111} (\bibinfo{year}{2001}),
  \eprint{arXiv:astro-ph/0105396}.

\bibitem[{\citenamefont{Manera et~al.}(2009)\citenamefont{Manera, Sheth, and
  Scoccimarro}}]{man09}
\bibinfo{author}{\bibfnamefont{M.}~\bibnamefont{Manera}},
  \bibinfo{author}{\bibfnamefont{R.~K.} \bibnamefont{Sheth}}, \bibnamefont{and}
  \bibinfo{author}{\bibfnamefont{R.}~\bibnamefont{Scoccimarro}}
  (\bibinfo{year}{2009}), \eprint{0906.1314}.

\bibitem[{\citenamefont{{Carlstrom} et~al.}(2002)\citenamefont{{Carlstrom},
  {Holder}, and {Reese}}}]{car02}
\bibinfo{author}{\bibfnamefont{J.~E.} \bibnamefont{{Carlstrom}}},
  \bibinfo{author}{\bibfnamefont{G.~P.} \bibnamefont{{Holder}}},
  \bibnamefont{and} \bibinfo{author}{\bibfnamefont{E.~D.}
  \bibnamefont{{Reese}}}, \bibinfo{journal}{\araa}
  \textbf{\bibinfo{volume}{40}}, \bibinfo{pages}{643} (\bibinfo{year}{2002}),
  \eprint{arXiv:astro-ph/0208192}.

\bibitem[{\citenamefont{{Koester} et~al.}(2007)\citenamefont{{Koester},
  {McKay}, {Annis}, {Wechsler}, {Evrard}, {Bleem}, {Becker}, {Johnston},
  {Sheldon}, {Nichol} et~al.}}]{koe07}
\bibinfo{author}{\bibfnamefont{B.~P.} \bibnamefont{{Koester}}},
  \bibinfo{author}{\bibfnamefont{T.~A.} \bibnamefont{{McKay}}},
  \bibinfo{author}{\bibfnamefont{J.}~\bibnamefont{{Annis}}},
  \bibinfo{author}{\bibfnamefont{R.~H.} \bibnamefont{{Wechsler}}},
  \bibinfo{author}{\bibfnamefont{A.}~\bibnamefont{{Evrard}}},
  \bibinfo{author}{\bibfnamefont{L.}~\bibnamefont{{Bleem}}},
  \bibinfo{author}{\bibfnamefont{M.}~\bibnamefont{{Becker}}},
  \bibinfo{author}{\bibfnamefont{D.}~\bibnamefont{{Johnston}}},
  \bibinfo{author}{\bibfnamefont{E.}~\bibnamefont{{Sheldon}}},
  \bibinfo{author}{\bibfnamefont{R.}~\bibnamefont{{Nichol}}},
  \bibnamefont{et~al.}, \bibinfo{journal}{\apj} \textbf{\bibinfo{volume}{660}},
  \bibinfo{pages}{239} (\bibinfo{year}{2007}), \eprint{arXiv:astro-ph/0701265}.

\bibitem[{\citenamefont{{Johnston} et~al.}(2007)\citenamefont{{Johnston},
  {Sheldon}, {Wechsler}, {Rozo}, {Koester}, {Frieman}, {McKay}, {Evrard},
  {Becker}, and {Annis}}}]{joh07}
\bibinfo{author}{\bibfnamefont{D.~E.} \bibnamefont{{Johnston}}},
  \bibinfo{author}{\bibfnamefont{E.~S.} \bibnamefont{{Sheldon}}},
  \bibinfo{author}{\bibfnamefont{R.~H.} \bibnamefont{{Wechsler}}},
  \bibinfo{author}{\bibfnamefont{E.}~\bibnamefont{{Rozo}}},
  \bibinfo{author}{\bibfnamefont{B.~P.} \bibnamefont{{Koester}}},
  \bibinfo{author}{\bibfnamefont{J.~A.} \bibnamefont{{Frieman}}},
  \bibinfo{author}{\bibfnamefont{T.~A.} \bibnamefont{{McKay}}},
  \bibinfo{author}{\bibfnamefont{A.~E.} \bibnamefont{{Evrard}}},
  \bibinfo{author}{\bibfnamefont{M.~R.} \bibnamefont{{Becker}}},
  \bibnamefont{and} \bibinfo{author}{\bibfnamefont{J.}~\bibnamefont{{Annis}}},
  \bibinfo{journal}{ArXiv e-prints} \textbf{\bibinfo{volume}{709}}
  (\bibinfo{year}{2007}), \eprint{0709.1159}.

\bibitem[{\citenamefont{{Komatsu} et~al.}(2008)\citenamefont{{Komatsu},
  {Dunkley}, {Nolta}, {Bennett}, {Gold}, {Hinshaw}, {Jarosik}, {Larson},
  {Limon}, {Page} et~al.}}]{kom08}
\bibinfo{author}{\bibfnamefont{E.}~\bibnamefont{{Komatsu}}},
  \bibinfo{author}{\bibfnamefont{J.}~\bibnamefont{{Dunkley}}},
  \bibinfo{author}{\bibfnamefont{M.~R.} \bibnamefont{{Nolta}}},
  \bibinfo{author}{\bibfnamefont{C.~L.} \bibnamefont{{Bennett}}},
  \bibinfo{author}{\bibfnamefont{B.}~\bibnamefont{{Gold}}},
  \bibinfo{author}{\bibfnamefont{G.}~\bibnamefont{{Hinshaw}}},
  \bibinfo{author}{\bibfnamefont{N.}~\bibnamefont{{Jarosik}}},
  \bibinfo{author}{\bibfnamefont{D.}~\bibnamefont{{Larson}}},
  \bibinfo{author}{\bibfnamefont{M.}~\bibnamefont{{Limon}}},
  \bibinfo{author}{\bibfnamefont{L.}~\bibnamefont{{Page}}},
  \bibnamefont{et~al.}, \bibinfo{journal}{ArXiv e-prints}
  (\bibinfo{year}{2008}), \eprint{0803.0547}.

\bibitem[{\citenamefont{{Seljak} and {Zaldarriaga}}(1996)}]{sel96}
\bibinfo{author}{\bibfnamefont{U.}~\bibnamefont{{Seljak}}} \bibnamefont{and}
  \bibinfo{author}{\bibfnamefont{M.}~\bibnamefont{{Zaldarriaga}}},
  \bibinfo{journal}{\apj} \textbf{\bibinfo{volume}{469}}, \bibinfo{pages}{437}
  (\bibinfo{year}{1996}), \eprint{arXiv:astro-ph/9603033}.

\bibitem[{\citenamefont{{Cohn}}(2006)}]{coh06}
\bibinfo{author}{\bibfnamefont{J.~D.} \bibnamefont{{Cohn}}},
  \bibinfo{journal}{New Astronomy} \textbf{\bibinfo{volume}{11}},
  \bibinfo{pages}{226} (\bibinfo{year}{2006}), \eprint{arXiv:astro-ph/0503285}.

\bibitem[{\citenamefont{Erickson et~al.}(2009)\citenamefont{Erickson, Evrard,
  and Cunha}}]{smi09}
\bibinfo{author}{\bibfnamefont{B.}~\bibnamefont{Erickson}},
  \bibinfo{author}{\bibfnamefont{A.~E.} \bibnamefont{Evrard}},
  \bibnamefont{and} \bibinfo{author}{\bibfnamefont{C.}~\bibnamefont{Cunha}},
  \bibinfo{journal}{in preparation}  (\bibinfo{year}{2009}).

\bibitem[{\citenamefont{{Wu} et~al.}(2009)\citenamefont{{Wu}, {Rozo}, and
  {Wechsler}}}]{wu09}
\bibinfo{author}{\bibfnamefont{H.-Y.} \bibnamefont{{Wu}}},
  \bibinfo{author}{\bibfnamefont{E.}~\bibnamefont{{Rozo}}}, \bibnamefont{and}
  \bibinfo{author}{\bibfnamefont{R.~H.} \bibnamefont{{Wechsler}}},
  \bibinfo{journal}{ArXiv e-prints}  (\bibinfo{year}{2009}),
  \eprint{0907.2690}.

\end{thebibliography}
\end{document}